\documentclass[graybox, nosecnum]{svmult}

\usepackage{mathptmx}       
\usepackage{helvet}         
\usepackage{courier}        
\usepackage{type1cm}        
\usepackage{makeidx}         
\usepackage{graphicx}        
\usepackage{multicol}        
\usepackage[bottom]{footmisc}
\usepackage{hyperref}        
\usepackage{soul}            

\usepackage{url}

\makeindex             

\begin{document}
\title*{Sub-barrier fusion reactions}
\author{K. Hagino}
\institute{K. Hagino \at 
Department of Physics, Kyoto University, Kyoto 606-8502, Japan
\email{hagino.kouichi.5m@kyoto-u.ac.jp}}

\maketitle
\abstract{
The concept of compound nucleus was proposed by Niels Bohr in 1936 to explain 
narrow resonances observed in scattering of a slow neutron off atomic nuclei. 
A compound nucleus is a metastable state with a long lifetime, in which all 
the degrees of freedom are in a sort of thermal equilibrium.  
Fusion reactions are defined as reactions to form such compound nucleus 
by merging two atomic nuclei. Here a short description of heavy-ion fusion 
reactions at energies close the Coulomb barrier is presented. This includes: 
(i) an overview of a fusion process, (ii) a strong interplay between nuclear 
structure and fusion, (iii) fusion and multi-dimensional/multi-particle 
quantum tunneling, 
and (iv) fusion for superheavy elements. 
}

\section{Introduction}

\subsection{A general introduction to heavy-ion fusion reactions} 

A fusion reaction is defined as a reaction to form a compound nucleus, the 
concept of which was originally proposed by Niels Bohr in 1936 \cite{bohr1936}. 
In the year earlier, Enrico Fermi performed experiments with slow neutrons  
and observed many narrow resonances in scattering cross sections. 
The width of those resonances was typically order of 
a few eV (see e.g., Ref. \cite{asghar1966}), which 
is much smaller than a typical nuclear scale of order of MeV. 
This implies that the resonances formed by reactions of slow neutrons are very 
long-lived. Bohr considered that the energy of the incident neutron was distributed 
among the other nucleons in a nucleus after many collisions, and a kind of thermal equilibrium state 
was formed. This is the concept of compound nucleus which Bohr proposed. 
Since a nucleus is a finite system, the energy may be 
concentrated once again to one of the 
neutrons in a nucleus, and that neutron is scattered off from a nucleus. 
This happens only at a long time after the compound nucleus is formed, leading to 
narrow resonance widths. 

\begin{figure}[tb]
\begin{center}
\includegraphics*[width=0.8\textwidth]{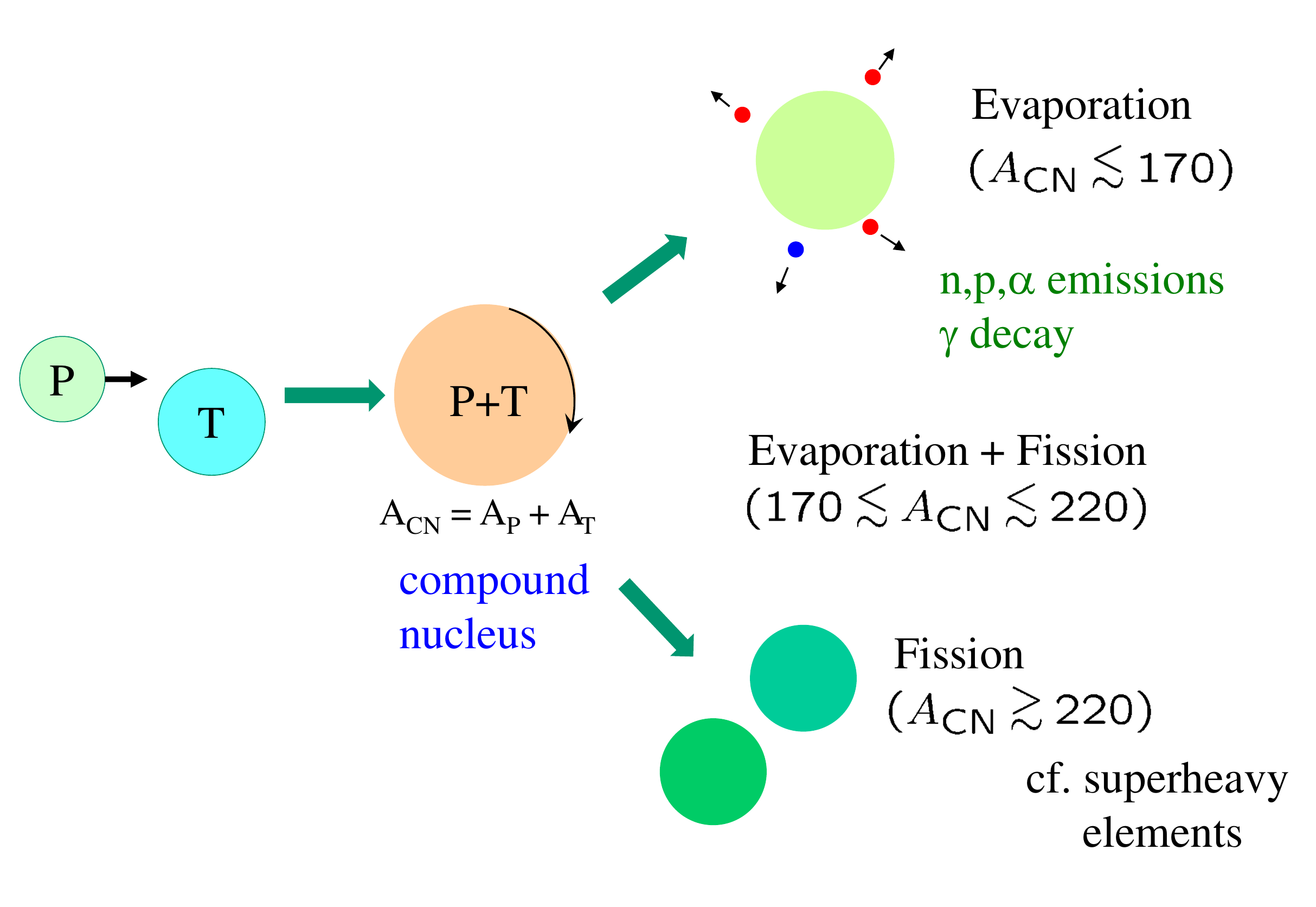}
\caption{
A schematic illustration of heavy-ion fusion reactions.}
\label{fig:overview}
\end{center}
\end{figure}

Similar compound nuclei are formed in heavy-ion fusion reactions by bombarding two 
heavy nuclei. 
Such fusion reaction plays an important role in several phenomena in nuclear physics, 
such as the energy production in stars, nucleosyntheses, and formations of superheavy 
elements. Theoretically, fusion, as well as fission, are large amplitudes motions of 
quantum many-body systems with a strong interaction, and their microscopic 
understanding is one of the ultimate goals of nuclear physics \cite{bender2020}. 

Figure \ref{fig:overview} shows a schematic illustration of a fusion 
process. At first a projectile nucleus (``P'') collides with a target nucleus 
(``T'') and forms a unified nucleus (``P+T''), i.e., a compound nucleus. 
Since the projectile nucleus brings the energy and the angular momentum into 
a system, the compound nucleus is at high excitation energies with large 
angular momenta. For light compound nucleus with the mass number less than 
about 170, the compound nucleus decays mainly by emitting neutrons, protons, 
alpha particles, and gamma rays. Such process is called an evaporation process. 
For heavy compound nucleus with the mass number larger than about 220fission 
dominates the decay process of the compound nucleus. This is particularly the case 
for fusion reactions relevant to superheavy nuclei. 
For compound nuclei with intermediate mass numbers, the evaporation and the fission 
processes compete with each other. 
In any case, fusion cross sections are measured by detecting 
residues of the evaporation process (called ``evaporation residues'') and/or 
fission fragments. 

\begin{figure}[tb]
\begin{center}
\includegraphics*[width=0.8\textwidth]{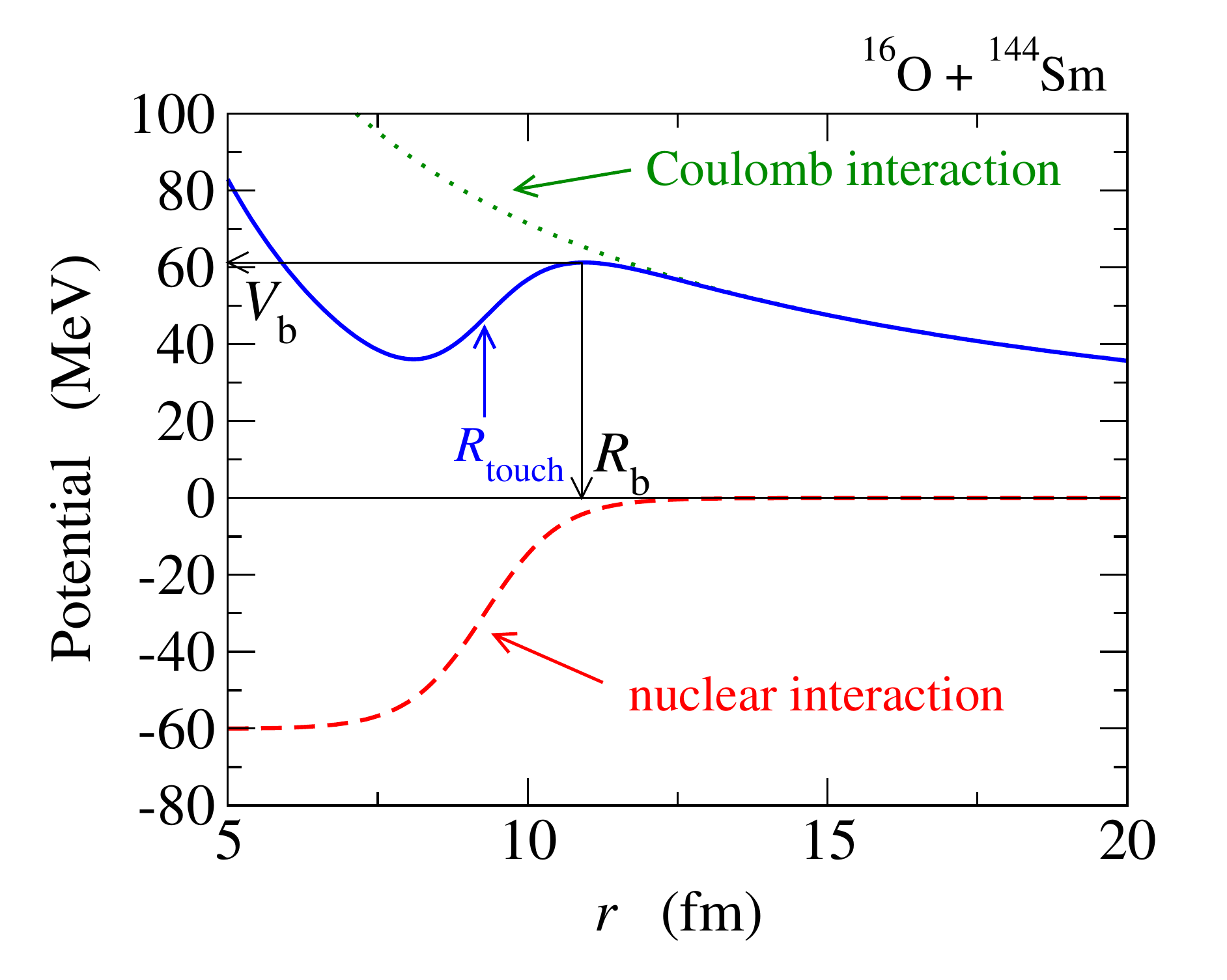}
\caption{
An internucleus potential between two nuclei as a function of 
the distance between them (the solid line). 
The $^{16}$O+$^{144}$Sm reaction is considered as a typical example. 
The Coulomb and the nuclear contributions are denoted by the dotted and the dashed 
lines, respectively. $R_b$ and $V_b$ denote the position and the height of the 
Coulomb barrier, respectively. $R_{\rm touch}$ is the distance at which two 
nuclei touch with each other. }
\label{fig:barrier}
\end{center}
\end{figure}

Figure \ref{fig:barrier} shows a typical potential between two nuclei 
as a function of the distance $r$ between them. 
Two different interactions are involved here. Firstly, a nucleus has a positive 
charge, and the Coulomb interaction acts between two nuclei. This is a long-range 
and repulsive interaction. When the distance between the two nuclei 
gets smaller, an attractive short range nuclear interaction (i.e., the strong 
interaction) becomes active. Because of the cancellation of these two, a potential 
barrier, referred to as the Coulomb barrier, is formed at some distance $R_b$, 
which is usually larger than the touching radius $R_{\rm touch}$ at which the two 
nuclei touch with each other. 
The height of the Coulomb barrier, $V_b$, specifies the energy scale of the 
reaction system. 

In this article, we overview the fusion dynamics at energies 
around the Coulomb barrier, that is, subbarrier fusion reactions. 
There are two obvious reasons 
why the subbarrier region is important. 
One is related to fusion 
reactions to form superheavy elements. Usually such experiments are carried 
out at energies slightly above the Coulomb barrier. For instance, in the 
$^{209}$Bi($^{70}$Zn,n)$^{278}$Nh reaction to form the element 113 (Nihonium), 
the experiments were performed at $E_{\rm c.m.}=$ 262 MeV in the center of mass 
frame \cite{morita2004,morita2007,morita2012}, 
while a barrier height for this reaction is around 260 MeV if the Bass 
potential \cite{bass1980} is employed. 
Fusion reactions for superheavy elements will be further discussed in a later section 
in this article. The second obvious reason to discuss the subbarrier region is 
in connection to nuclear astrophysical reactions. Nuclear fusion reactions in 
stars, such as the $^{12}$C+$^{12}$C reaction, take place at extremely low energies, 
for which direct measurements of fusion cross sections are difficult. 
One thus has to extrapolate measured fusion cross sections at higher energies 
down to the energy region 
relevant to nuclear astrophysics. In order to do reliable extrapolations, deep 
understandings of the fusion dynamics in the subbarrier region are crucially 
important. 

Besides these two obvious reasons, the reaction dynamics of subbarier fusion 
is intriguing in its own. Firstly, it is known that nuclear structure affects 
significantly nuclear fusion, and thus there is a strong interplay between nuclear 
structure and nuclear reaction there. This is in contrast to high energy nuclear 
reactions, at which the reaction dynamics is much simpler. Secondly, subbarrier fusion 
reactions can be regarded as a typical example of many-particle tunneling phenomena. 
In order for fusion to take place, two nuclei have to get close at least 
to the touching radius, and thus fusion occurs only by quantum tunneling effect when the 
incident energy is below the Coulomb barrier (see Fig. \ref{fig:barrier}). 
An interesting fact in atomic nuclei 
is that there are many types of intrinsic degrees of freedom 
which can affect quantum tunneling, that is, there are several types of collective 
vibrations as well as nuclear deformation with several multipolarities. 
Moreover, the energy dependence of the tunneling rate can be studied in fusion 
reactions by varying the incident energy, 
in a marked contrast to other tunneling phenomena in nuclear physics, such 
as alpha decays, for which the energy is basically fixed by the decay $Q$-value. 
Heavy-ion fusion reactions can be thus considered as an ideal playground to study 
many-particle quantum tunneling with many degrees of freedom. 

\subsection{Earlier review articles and textbooks} 

A few review articles have been published on the subject of subbarrier fusion 
reactions. While Refs. \cite{BT98,hagino2012} discuss theoretical aspects of 
subbarrier fusion reactions, Ref. \cite{DHRS98} summarizes experimental observations 
in subbarrier fusion reactions 
from a viewpoint of the so called fusion barrier distributions. 
Refs. \cite{Back14,MS17,jiang2021} discuss heavy-ion fusion reactions at 
deep subbarrier energies, at which fusion cross sections appear to 
be hindered as compared to simple extrapolations of fusion cross sections 
at subbarrier energies. Subbarrier fusion reactions are discussed also in 
many textbooks of nuclear reactions, see e.g. 
Refs. \cite{frobrich1996,bertulani2004,thompson2009,canto2013}. 

\section{Potential Model}

\subsection{Potential model and the Wong formula}

The simplest approach to fusion reactions is to employ the potential model, 
in which one considers inert projectile and target nuclei and 
assumes some potential between them. 
For fusion reactions of medium-heavy nuclei, 
it is considered to be a good approximation to assume that a compound nucleus is formed 
automatically once the touching position is achieved. 
Fusion cross sections $\sigma_{\rm fus}(E)$ are then given by 
\begin{equation}
\sigma_{\rm fus}(E)=\frac{\pi}{k^2}\sum_l(2l+1)P_l(E), 
\label{eq:crosssections0}
\end{equation}
where $E$ is the bombarding energy in the center of mass frame 
and $k=\sqrt{2\mu E/\hbar^2}$ is the wave number for the relative motion between 
the two nuclei with the reduced mass $\mu$. $l$ is the orbital angular momentum 
for the relative motion, 
and $P_l(E)$ 
is the probability to reach the touching configuration. 
The factor $2l+1$ is simply a statistical weight coming from the fact that 
the probability $P_l$ does not depend on the $z$-component of $l$. 
Notice that $P_l(E)$ is nothing but the penetrability of the Coulomb barrier. 
This can be evaluated e.g., by adding a short range 
absorbing potential to an internucleus potential. 
Such absorbing potential in general describes any process besides elastic 
scattering, but to a good approximation it 
simulates the compound nucleus formation as long as it 
is well confined inside the Coulomb barrier. 

Based on this approach, Wong has derived a simple compact formula for fusion 
cross sections \cite{Wong1973} (see also Ref. \cite{RH2015}). To this end, 
he first approximated the Coulomb barrier by an inverted parabola, 
\begin{equation}
V(r)\sim V_b-\frac{1}{2}\mu\Omega^2(r-R_b)^2, 
\label{eq:parabolic}
\end{equation}
for which the penetrability can be given analytically as 
\begin{equation}
P_0(E)=\frac{1}{1+\exp\left[\frac{2\pi}{\hbar\Omega}(V_b-E)\right]}.
\label{eq:P0swave}
\end{equation}
For non-zero partial waves, he considered $l$-independent barrier position 
and curvature, and replaced $P_l(E)$ with 
\begin{equation}
P_l(E)\sim P_0\left(E-\frac{l(l+1)\hbar^2}{2\mu R_b^2}\right). 
\end{equation}
Finally, Wong replaced the sum in Eq. (\ref{eq:crosssections0}) by an integral 
\begin{equation}
\sigma_{\rm fus}(E)=\frac{\pi}{k^2}\sum_l(2l+1)P_l(E) 
\to \frac{\pi}{k^2}\int dl\,(2l+1)P(l,E)
\end{equation}
to obtain the so called Wong formula given by, 
\begin{equation}
\sigma_{\rm fus}(E)
=\frac{\hbar\Omega}{2E}\,R_b^2\,\ln\left[
1+\exp\left(\frac{2\pi}{\hbar\Omega}(E-V_b)\right)\right]. 
\label{eq:wong}
\end{equation}
Notice that the first energy derivative of $E\sigma_{\rm fus}$ from this 
formula is proportional to the $s$-wave penetrability of the Coulomb barrier, 
\begin{equation}
\frac{d}{dE}[E\sigma_{\rm fus}(E)]
=\pi R_b^2\,P_0(E),
\label{eq:bd0}
\end{equation}
where $P_0(E)$ is given by Eq. (\ref{eq:P0swave}). 

\begin{figure}[tb]
\begin{center}
\includegraphics*[width=1.0\textwidth]{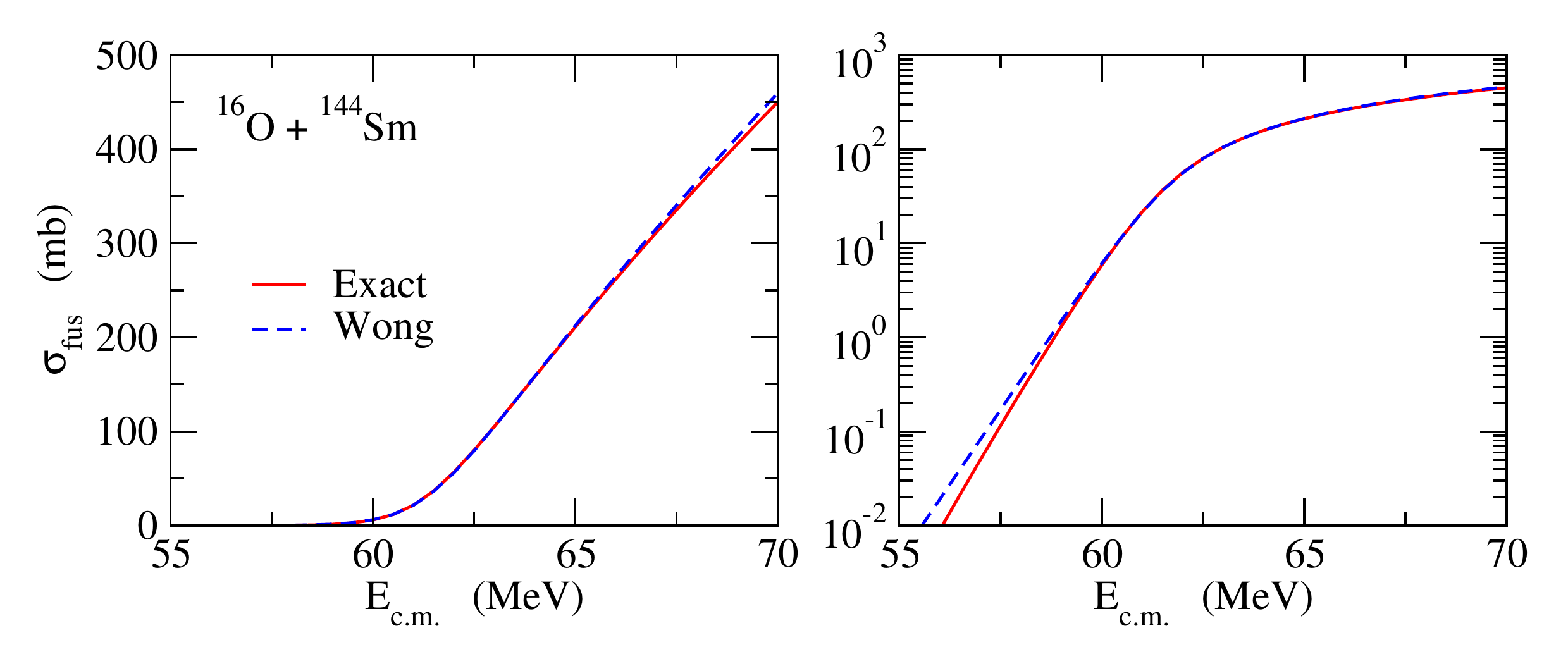}
\caption{
A comparison of the fusion cross sections obtained with the Wong 
formula (the dashed lines) with the 
exact fusion cross sections (the solid lines) 
for the $^{16}$O+$^{144}$Sm system. 
The left and the right panels show the results with the linear and the 
logarithmic 
scales, respectively. 
}
\label{fig:wong}
\end{center}
\end{figure}

Figure \ref{fig:wong} shows calculated fusion cross sections 
for the $^{16}$O+$^{144}$Sm system. The dashed lines are obtained with the 
Wong formula, while the solid lines are obtained by numerically 
solving the Schr\"odinger 
equation for each partial wave to obtain $P_l(E)$. The left and the right 
panels show the results in the linear and the logarithmic scales, respectively. 
One can see that the Wong formula works well, except for the region well below the 
Coulomb barrier (the height of the Coulomb barrier is about $V_b=61.25$ MeV 
for this system, see Fig. \ref{fig:barrier}), at which the Wong formula overestimates 
the fusion cross sections. The overestimate of fusion cross sections at energies 
well below the barrier is because the parabolic approximation 
(\ref{eq:parabolic}) used in the Wong formula 
underestimates the width of the potential barrier, which results in the overestimate of the penetrabilities. 

\subsection{Comparisons with experimental data}

\begin{figure}[tb]
\begin{center}
\includegraphics*[width=1.0\textwidth]{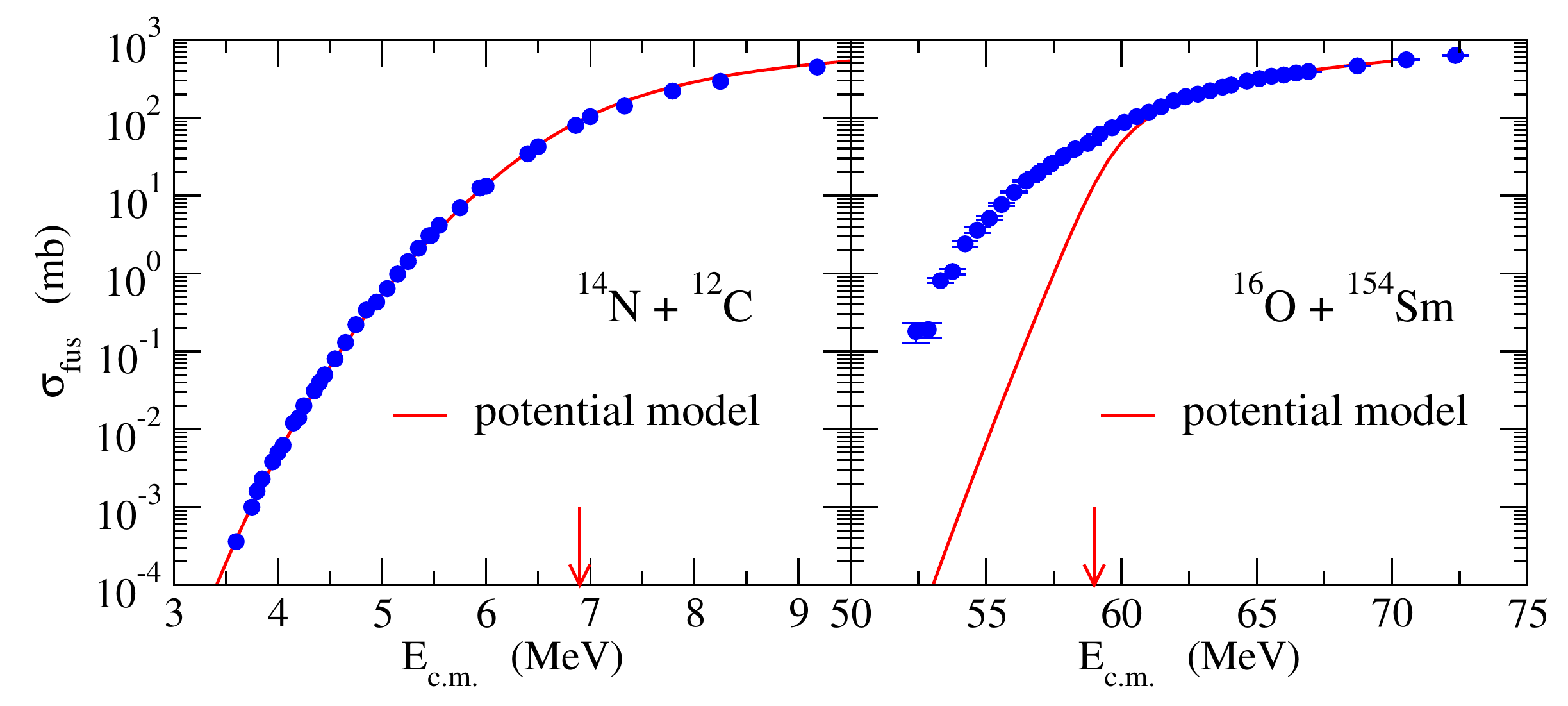}
\caption{
Fusion cross sections for 
the $^{14}$N+$^{12}$C system (the left panel) 
and for the $^{16}$O+$^{154}$Sm system (the right panel) obtained with the 
potential model. The arrows indicate the height of the Coulomb barrier for 
each system. The experimental data are taken from Refs. 
\cite{switkowski1977,Leigh1995}. 
}
\label{fig:comparison}
\end{center}
\end{figure}

Figure \ref{fig:comparison} compares fusion cross sections obtained with the potential 
model to the experimental data for the 
$^{14}$N+$^{12}$C (the left panel) 
and the $^{16}$O+$^{154}$Sm (the right panel) systems. 
The height of the Coulomb barrier is around $V_b\sim$ 6.9 MeV 
for the $^{14}$N+$^{12}$C system and 
$V_b\sim$ 59 MeV for the $^{16}$O+$^{154}$Sm system. For the 
$^{14}$N+$^{12}$C system, one can see that the potential model works well. 
On the other hand, for the $^{16}$O+$^{154}$Sm system, the potential model 
largely underestimates the fusion cross sections at energies below the 
Coulomb barrier, even though it works well at energies above the barrier. 
This phenomenon is referred to as the subbarrier enhancement of fusion cross 
sections, and has been systematically observed in a number of 
systems \cite{beckerman1985,beckerman1988}. 

\section{Fusion of deformed nuclei}

\begin{figure}[tb]
\begin{center}
\includegraphics*[width=0.4\textwidth]{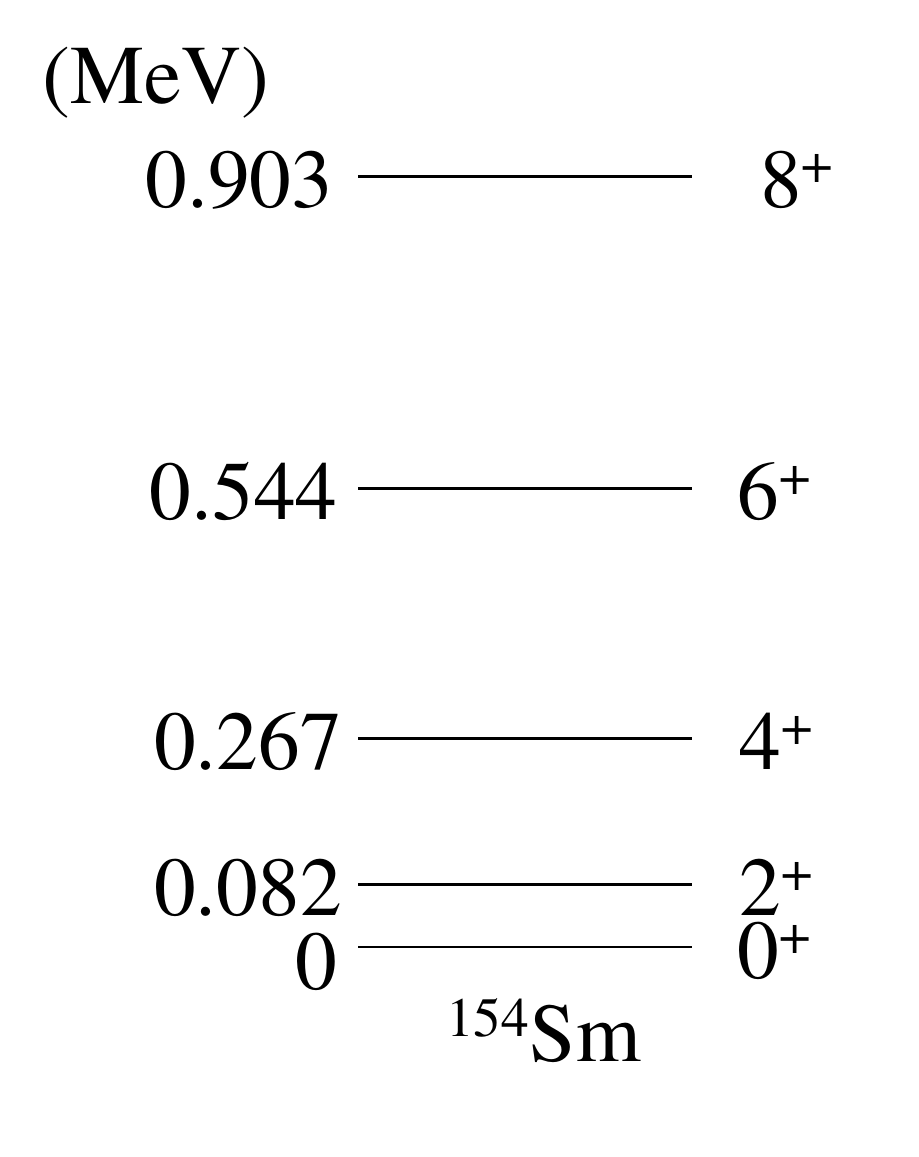}
\includegraphics*[width=0.5\textwidth]{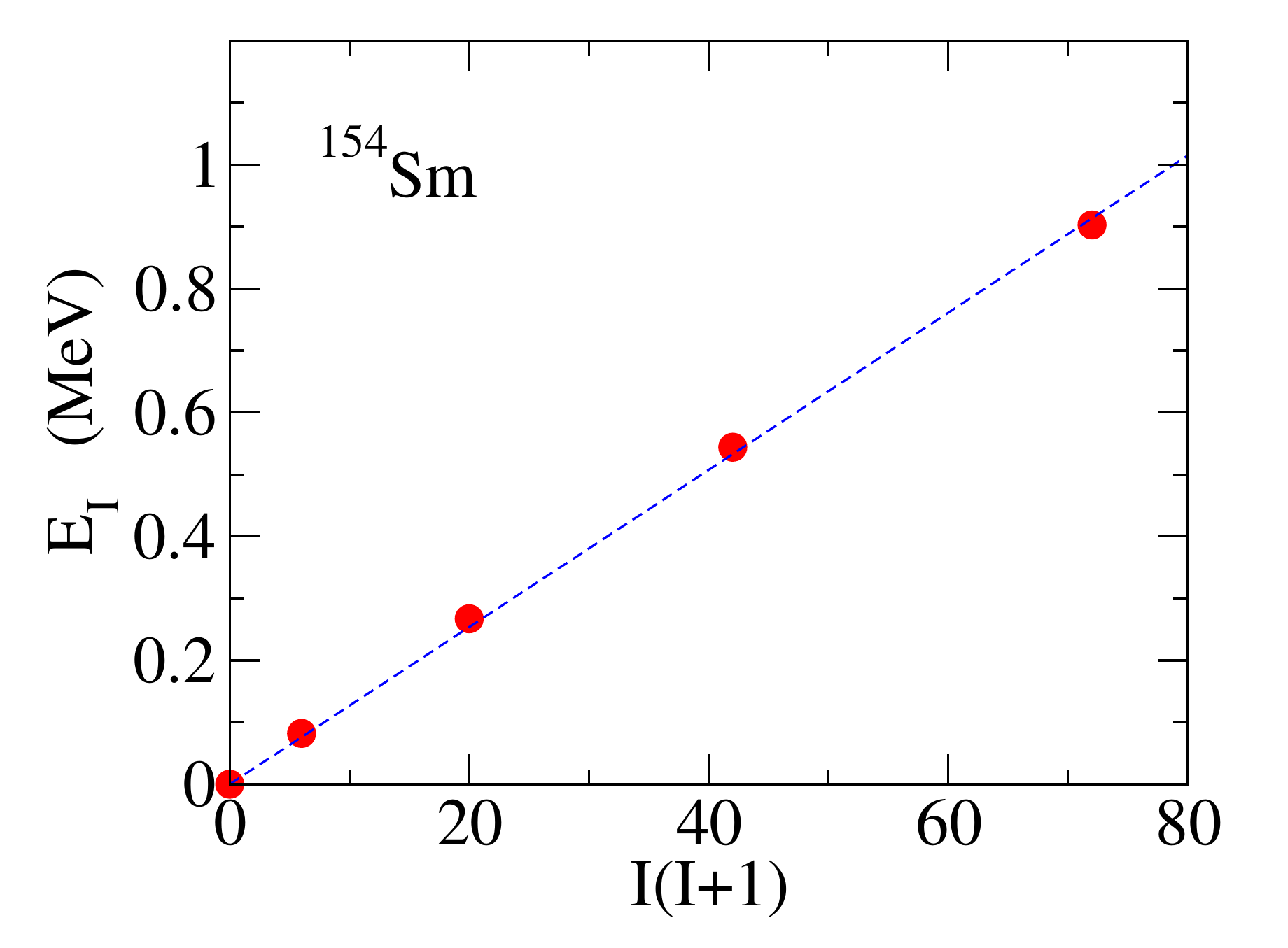}
\caption{
The spectrum of the $^{154}$Sm nucleus. Each level is 
specified by its angular momentum $I$ and parity $\pi$ as $I^\pi$. 
}
\label{fig:154sm}
\end{center}
\end{figure}

\begin{figure}[tb]
\begin{center}
\includegraphics*[width=0.6\textwidth]{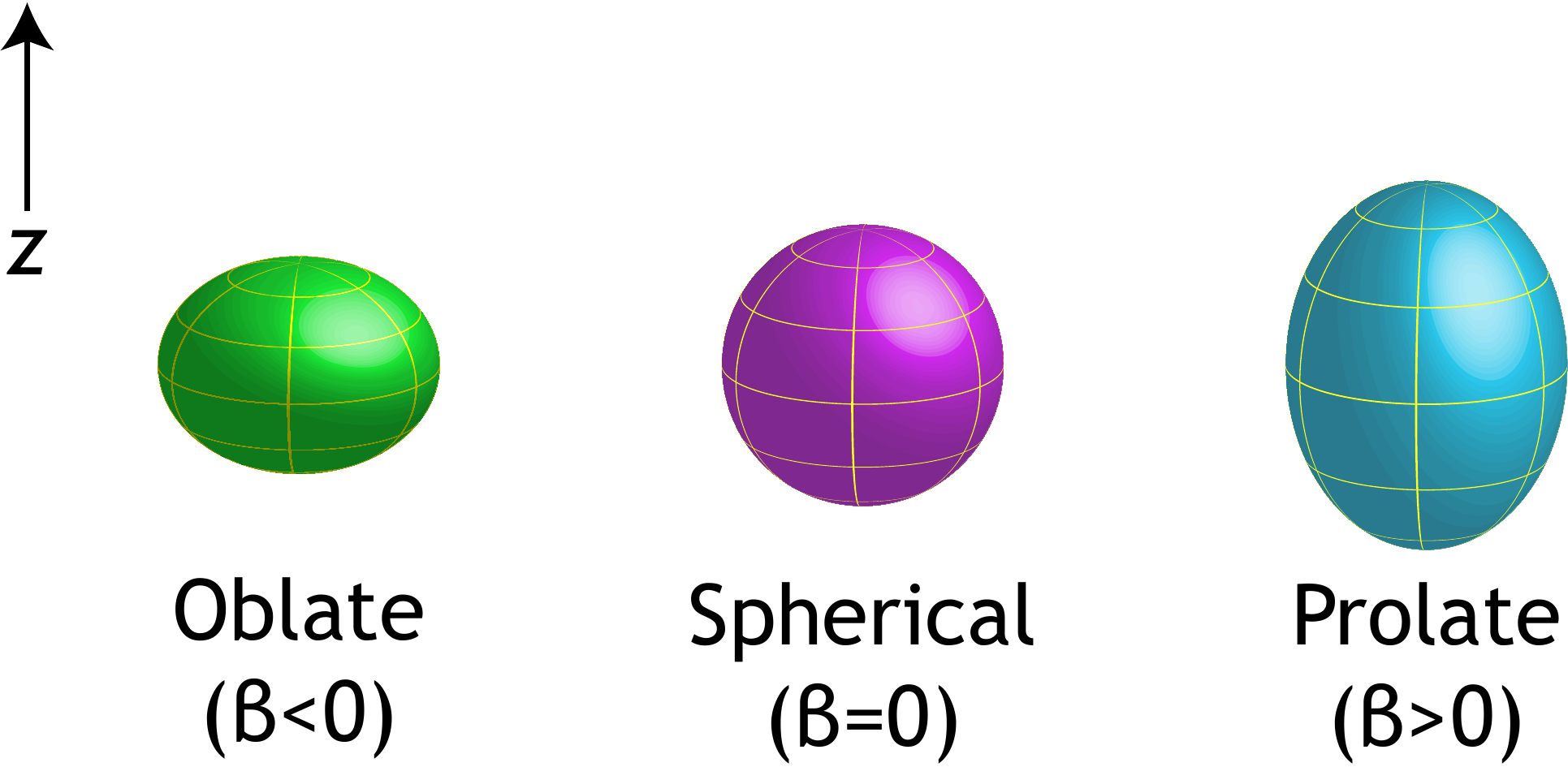}
\includegraphics*[width=0.4\textwidth]{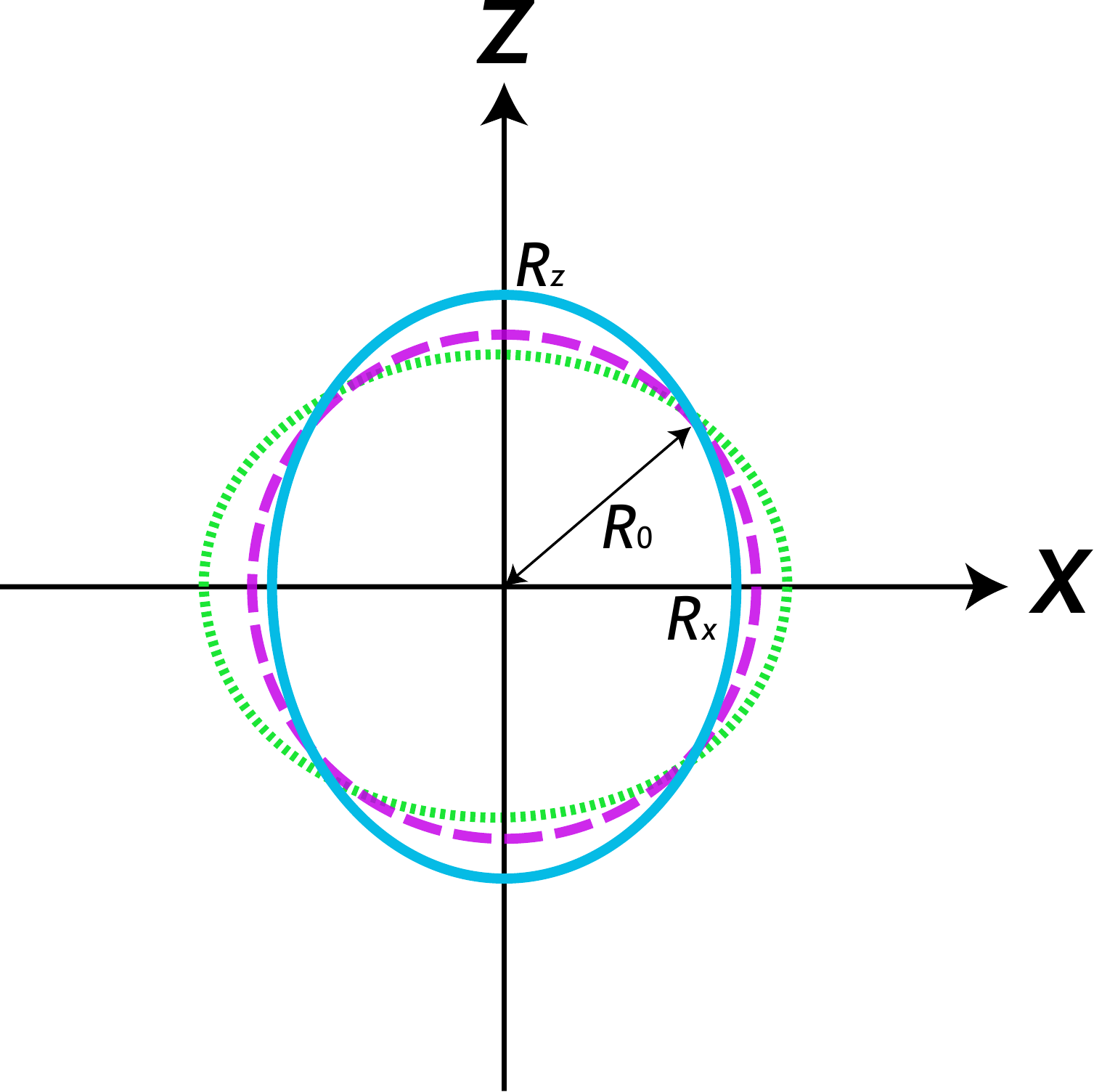}
\caption{
A schematic view of nuclear deformation, see also Ref. \cite{Naito2021}.
}
\label{fig:deformation}
\end{center}
\end{figure}

The subbarrier enhancement of fusion cross sections 
for the $^{16}$O+$^{154}$Sm system 
shown in Fig. \ref{fig:comparison}
can be easily explained if one notices that the $^{154}$Sm nucleus is a typical deformed nucleus. 
This nucleus exhibits characteristic rotational excitations, for which 
the energy of a state with the angular momentum $I$ is proportional to 
$I(I+1)$ (see Fig. \ref{fig:154sm}). This is interpreted 
as that $^{154}$Sm is statically deformed in the ground 
state (see Fig. \ref{fig:deformation}). 
For axially symmetric shape, the nuclear deformation is often 
characterized by the deformation parameters $\{\beta_\lambda\}$ defined as 
\begin{equation}
R(\theta)=R_0\left(1+\sum_\lambda \beta_\lambda Y_{\lambda 0}(\theta)\right),
\end{equation}
where $R(\theta)$ is the angle-dependent radius of a nucleus, $R_0$ is the 
radius in the spherical limit, $\theta$ is the 
angle measured from the symmetric axis, and $Y_{\lambda 0}(\theta)$ is the 
spherical harmonics (see the lower figure of Fig. \ref{fig:deformation}). 
For the $^{154}$Sm nucleus, the deformation parameters are 
$\beta_2\sim 0.30$ for $\lambda$=2 (the quadrupole deformation) 
and 
$\beta_4\sim 0.05$ for $\lambda$=4 (the hexadecapole deformation) \cite{Leigh1995,leigh1993}. 

\begin{figure}[tb]
\begin{center}
\includegraphics*[width=1.0\textwidth]{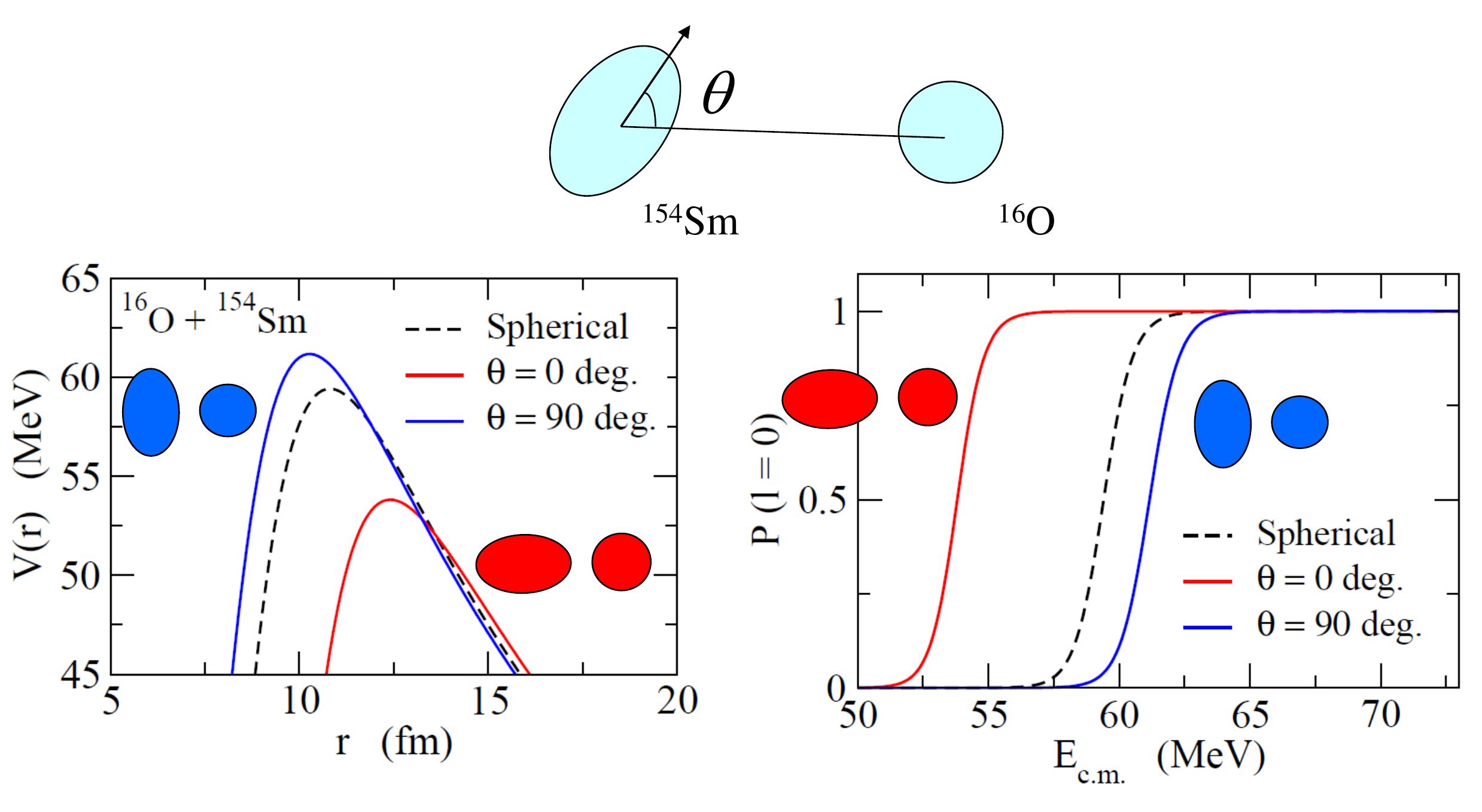}
\caption{
The angle-dependent internucleus potential (the left panel)  
for the $^{16}$O+$^{154}$Sm system, for which the orientation angle of 
the deformed $^{154}$Sm is denoted by $\theta$. 
The corresponding penetrabilities for $s$-wave scattering are 
shown in the right panel. 
The dashed lines show the potential (the left panel) 
and the penetrability (the right panel) in the 
spherical limit. 
}
\label{fig:defpot2}
\end{center}
\end{figure}

When a target nucleus is deformed, the internucleus potential depends on the 
orientation angle of the deformed nucleus. 
When the projectile nucleus approaches the 
target nucleus from the direction of the longer axis of the target, 
the Coulomb barrier is lowered as compared to the potential in the spherical 
limit. This is because the nuclear attraction acts from longer distances. 
The lowering of a barrier results in an enhancement of 
the penetrability. 
For a prolately deformed nucleus with $\beta_2>0$, this corresponds to the 
angle $\theta=0$ (see Fig. \ref{fig:defpot2}). 
The opposite happens when 
the projectile approaches from the direction of the shorter axis of the target. 
For a prolately deformed nucleus with $\beta_2>0$, this corresponds to the 
angle $\theta=\pi/2$. 

\begin{figure}[tb]
\begin{center}
\includegraphics*[width=0.495\textwidth]{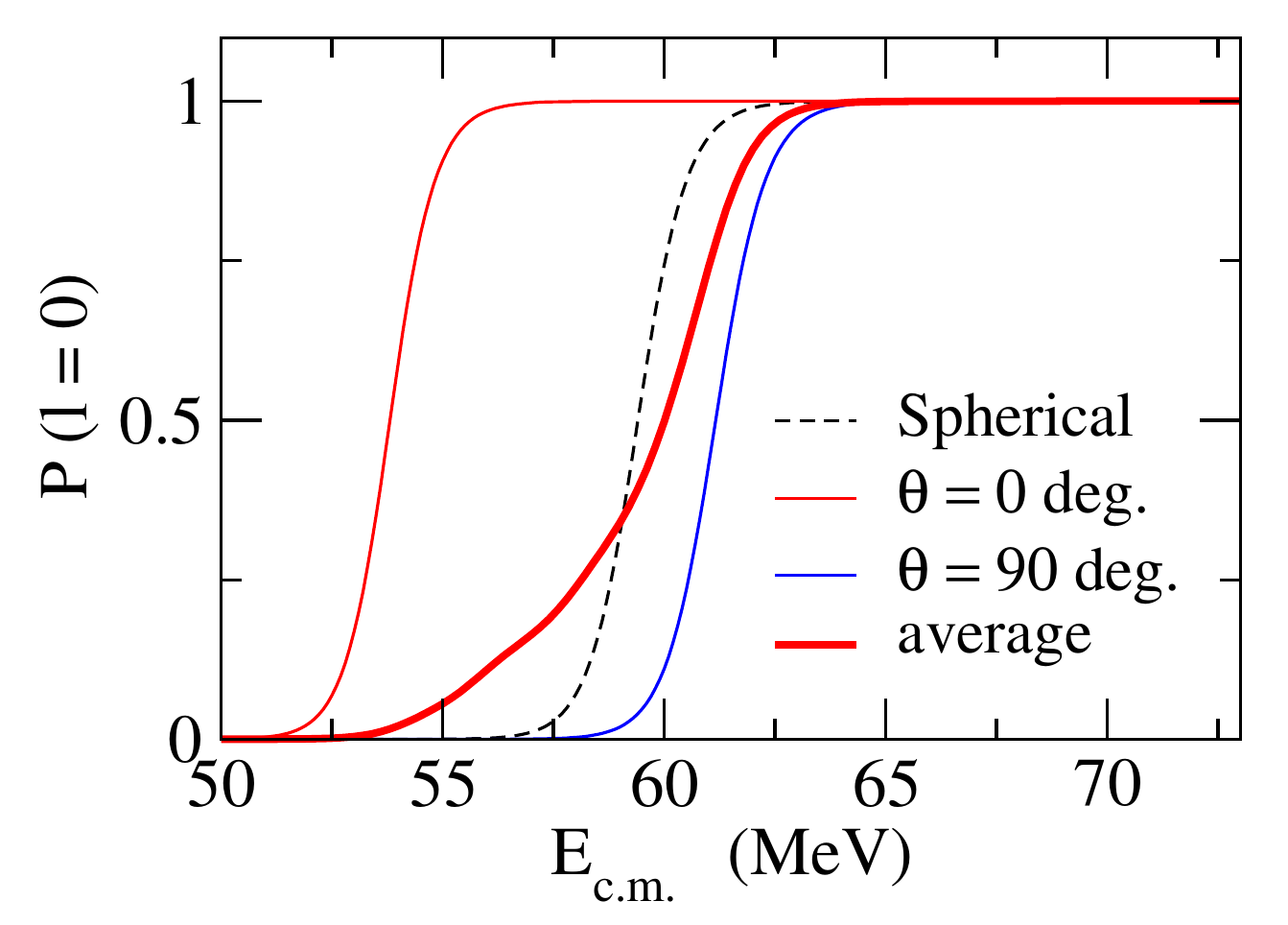}
\includegraphics*[width=0.495\textwidth]{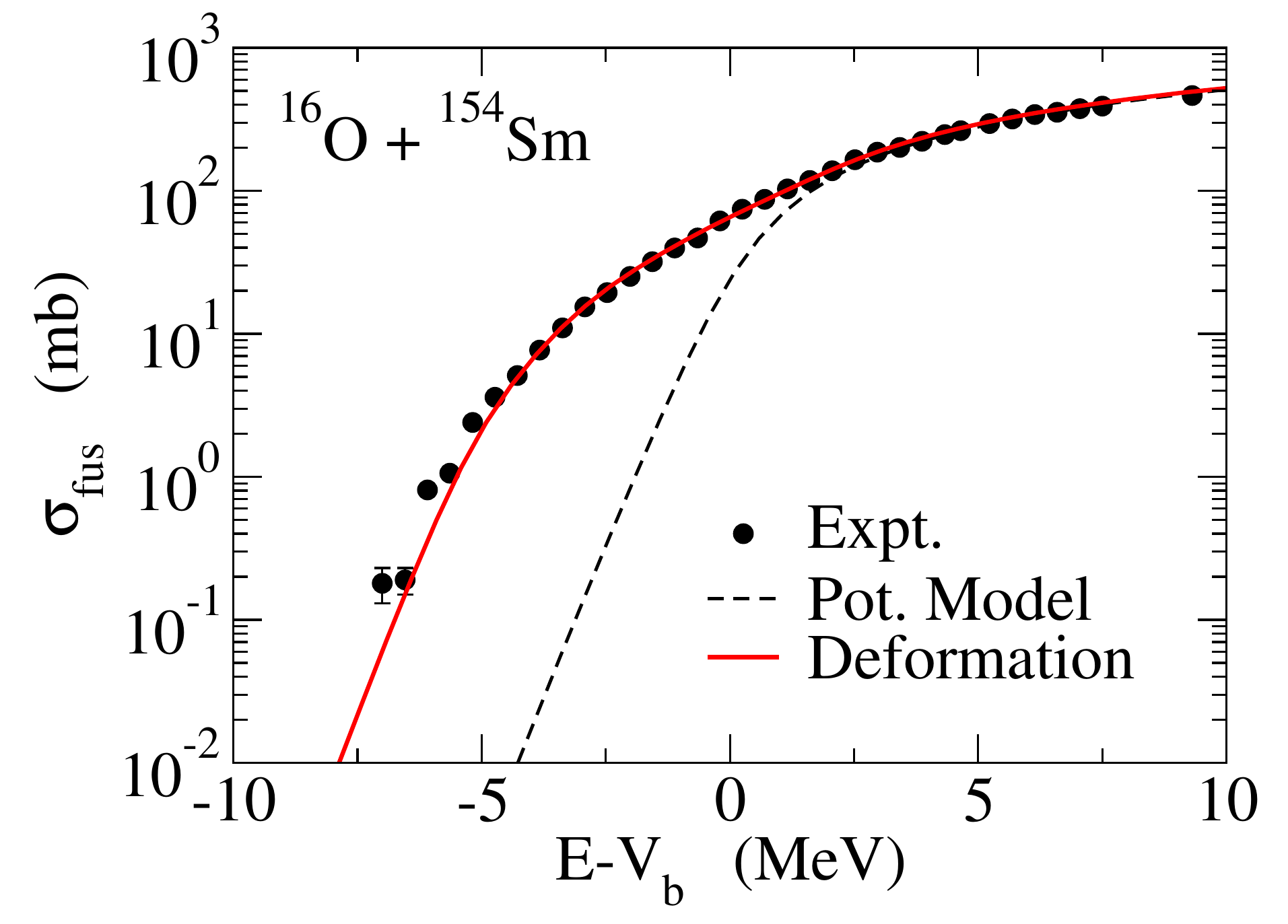}
\caption{
(The left panel)
The same as the right panel of 
Fig. \ref{fig:defpot2}, but with the penetrability 
obtained by averaging 
all the orientation angles (the thick solid line). 
(The right panel) 
The same as the right panel of Fig. \ref{fig:comparison}, but with 
the fusion cross sections 
obtained by averaging 
all the orientation angles (the solid line). 
The fusion cross sections are plotted as a function of energy relative to 
the height of the Coulomb barrier. 
}
\label{fig:defpot3}
\end{center}
\end{figure}

The total penetrability is computed by averaging the angle-dependent penetrability 
as 
\begin{equation}
P(E)=\int^1_0d(\cos\theta)\,P_0(E;\theta), 
\label{eq:defP}
\end{equation}
where $P_0(E;\theta)$ is the penetrability for the orientation angle $\theta$. 
The thick solid line in the left panel of Fig. \ref{fig:defpot3} is obtained 
in this way. 
Since the tunneling probability has an exponential dependence on the energy, 
the enhancement of the penetrability due to $\theta\sim 0$ leads to 
the main contribution 
to the total penetrability at energies below the barrier. 
The total penetrability is thus enhanced at these energies as compared 
to the penetrabilities in the spherical limit. 
This is the main mechanism for the subbarrier fusion cross sections 
shown in Fig. \ref{fig:comparison}. 
The formula (\ref{eq:defP}) can be actually extended to fusion cross 
sections $\sigma_{\rm fus}$ as, 
\begin{equation}
\sigma_{\rm fus}(E)=\int^1_0d(\cos\theta)\,\sigma^{(0)}_{\rm fus}(E;\theta). 
\label{eq:deffus-cross}
\end{equation}
The solid line in the right panel of 
Fig. \ref{fig:defpot3} is obtained in this way. One can see that 
the subbarrier enhancement of fusion cross sections for this system is well accounted 
for by taking into account the deformation of the $^{154}$Sm nucleus. 
This clearly demonstrates that 
there is a strong interplay between nuclear structure and 
heavy-ion fusion reactions at subbarrier energies.

\section{Coupled-channels approach}

\begin{figure}[tb]
\begin{center}
\includegraphics*[width=0.9\textwidth]{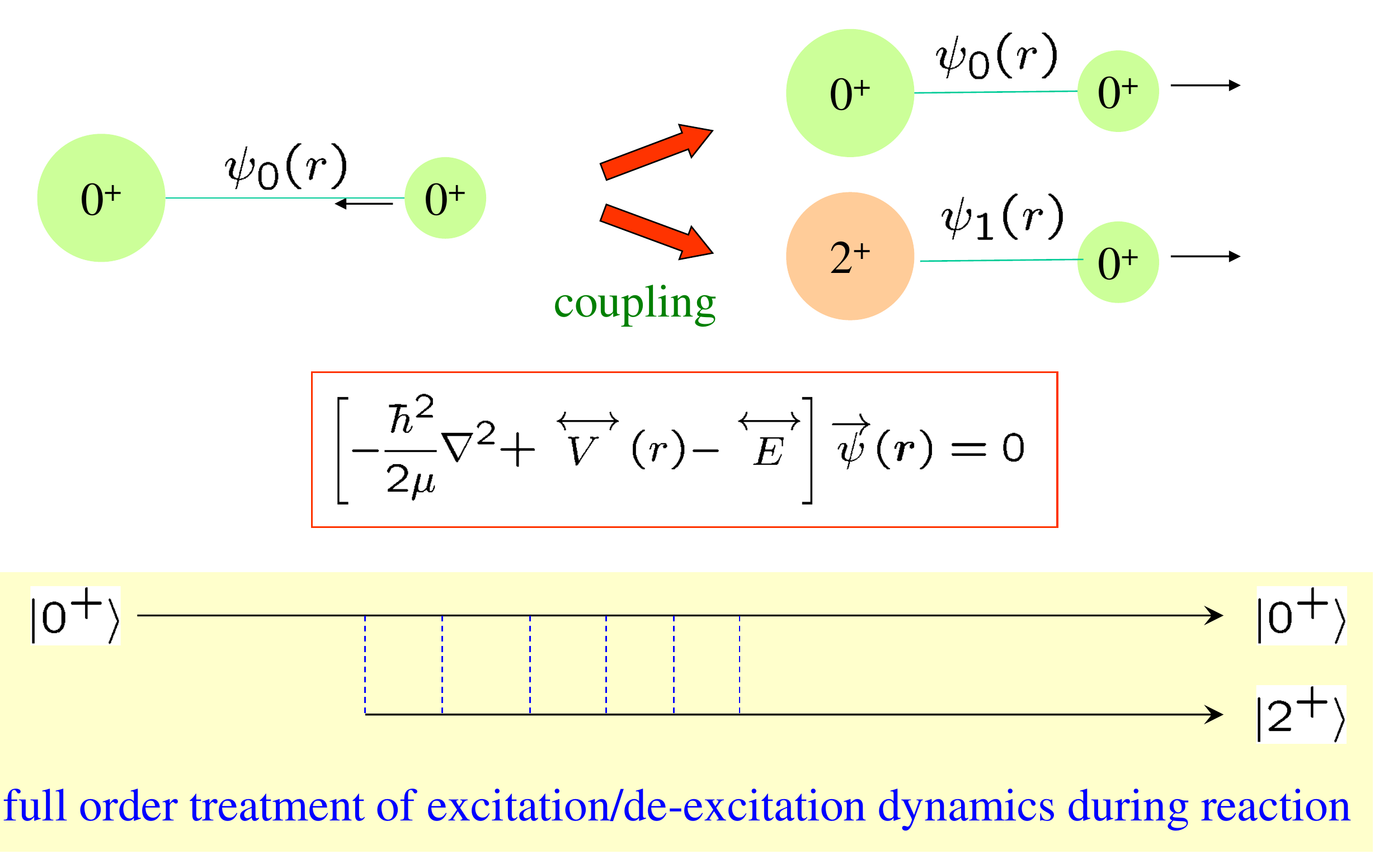}
\caption{
A schematic illustration of the coupled-channels approach. 
Here, excitations to a state with $I^\pi=2^+$ in a target nucleus 
is considered. 
$\vec{\psi}(r)$ is the wave function for the relative motion, which depends on 
the target states. 
}
\label{fig:CC}
\end{center}
\end{figure}

The subbarrier fusion enhancement discussed in the previous section 
has been observed also in systems with non-deformed target nuclei. 
It has been understood by now that the subbarrier fusion enhancement 
is caused by couplings of the relative motion between colliding nuclei 
to several low-lying collective excitations in the nuclei as well as 
particle transfer processes \cite{BT98,hagino2012,DHRS98}. 
The deformation effect discussed in the previous section is a special case, 
in which the rotational excitations due to a nuclear deformation can be 
taken into account in terms of orientation-dependent internucleus 
potential \cite{hagino2012}. 
In order to take into account such coupling effects, the coupled-channels 
approach has been developed \cite{hagino2012,hagino2021,tamura1965,satchler1983,broglia2004}. 
This is a quantal reaction theory schematically illustrated in 
Fig. \ref{fig:CC}. 
In this figure, couplings of the relative motion to a state in the target 
nucleus with the angular 
momentum $I=2$ and positive parity $\pi=+$ are considered. 
At the initial stage of the reaction, the target nucleus is in the ground state, 
$I^\pi=0^+$. The wave function for the relative motion for this configuration 
is denoted by $\psi_0(r)$. During the reactions, due to the interaction between the 
projectile and the target nuclei, the target nucleus may be 
excited to the $I^\pi=2^+$ state, and at the same time, the relative wave function 
is changed to $\psi_1(r)$. The coupling is taken into account by the off-diagonal 
components of the potential $V(r)$. The 
$I^\pi=0^+$ state may be de-excited to the 0$^+$ state, and thus 
the two wave functions $\psi_0(r)$ and $\psi_1(r)$ are coupled to each other. 
One then solves  in a non-perturbative manner coupled Schro\"odinger equations, which are referred to as 
coupled-channels equations, 
to determine the $S$-matrix, from which several reaction 
observables can be constructed. 

\begin{figure}[tb]
\begin{center}
\includegraphics*[width=0.7\textwidth]{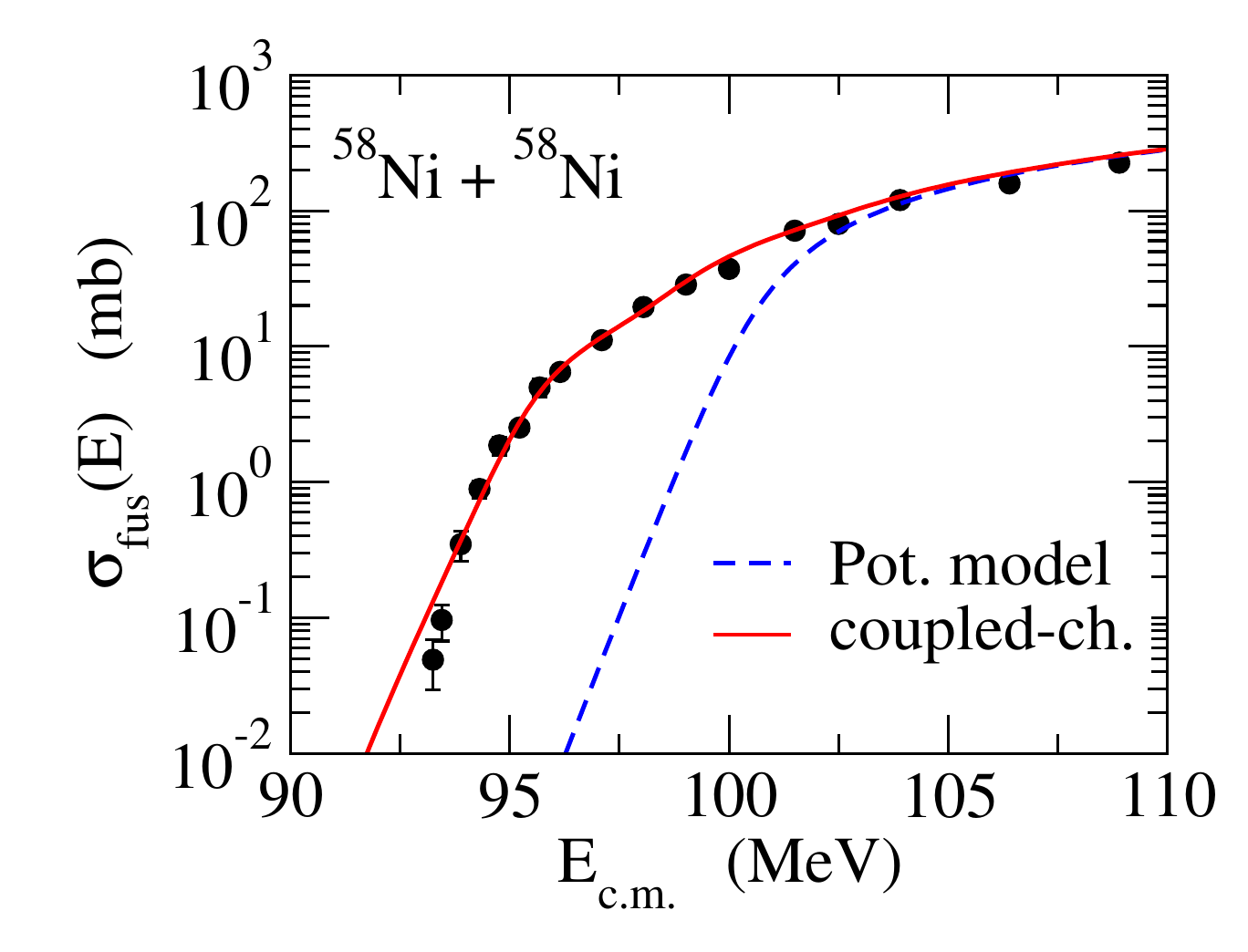}
\caption{Fusion cross sections for the $^{58}$Ni+$^{58}$Ni system. 
The dashed line shows the results of the potential model. 
The solid line shows the results of the coupled-channels 
calculations, in which 
the quadrupole excitations up to the double phonon states 
in each $^{58}$Ni 
are taken into account. The experimental data are 
taken from Ref. \cite{beckerman1981}.
}
\label{fig:58ni58ni}
\end{center}
\end{figure}

A few computer codes are available for coupled-channels 
calculations, such as {\tt ECIS} \cite{ecis,lepine-szily2021}, 
{\tt FRESCO} \cite{thompson2009,fresco}, and {\tt CCFULL}\cite{ccfull}. 
As an example, Fig. \ref{fig:58ni58ni} shows fusion cross sections for the 
$^{58}$Ni+$^{58}$Ni system calculated with the code {\tt CCFULL}. 
Here, the quadrupole excitations up to the double phonon states 
are taken into account in each of the $^{58}$Ni nucleus. 
By taking into account the excitations of the $^{58}$Ni nuclei, 
the subbarrier enhancement of fusion cross sections is well reproduced. 
One may regard this as a clear example of {\it coupling assisted tunneling} 
phenomenon. 

\begin{figure}[tb]
\begin{center}
\includegraphics*[width=0.6\textwidth]{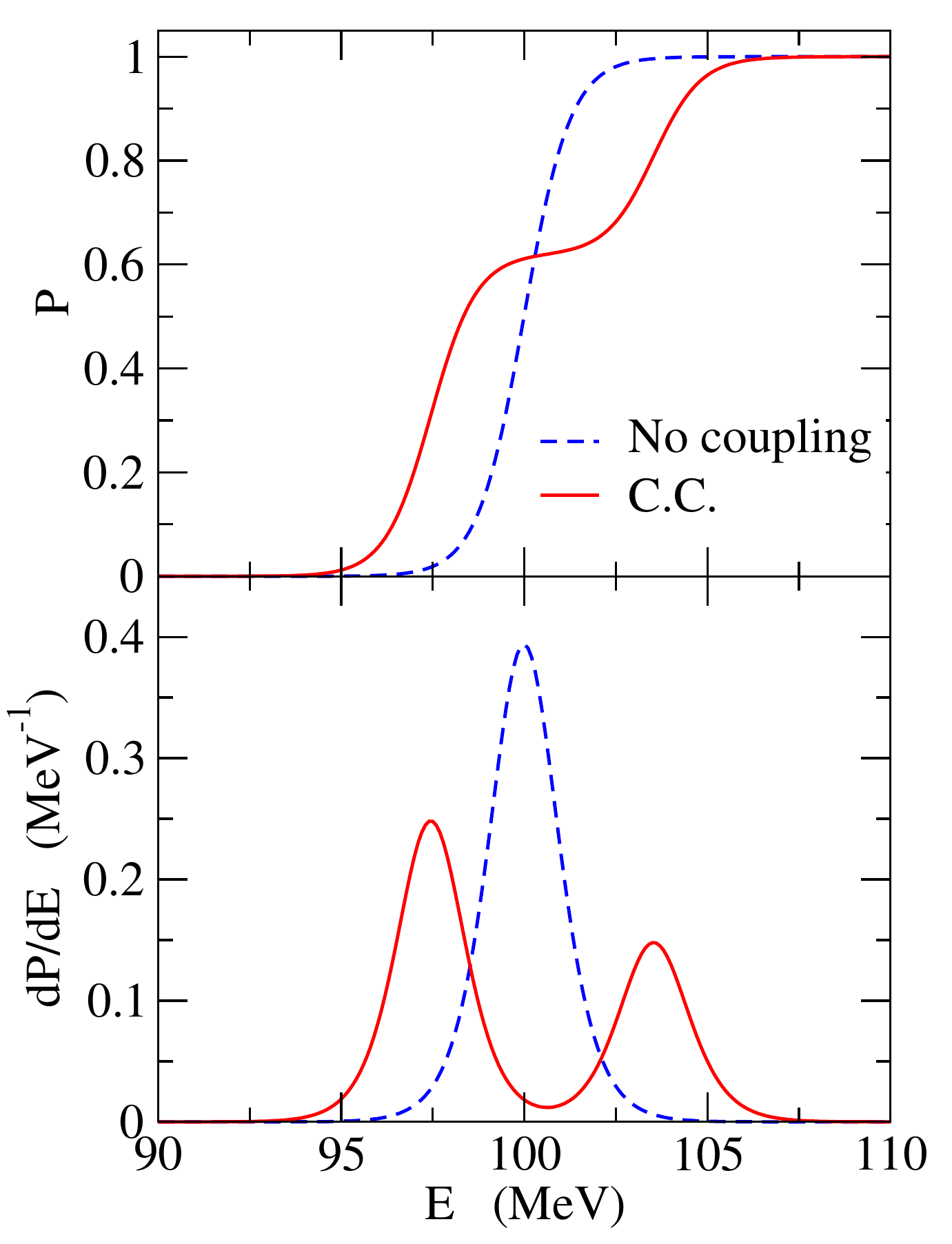}
\caption{
(The upper panel)
The penetrability of a one-dimensional 
Gaussian barrier in the presence of the channel coupling effects. 
The height of the barrier is set to be 100 MeV. 
The dotted line denotes the result without the channel coupling, while 
the solid line shows the result of the 
coupled-channels calculations.
(The lower panel)
The energy derivative of the penetrability shown in the upper panel. 
}
\label{fig:cc2ch}
\end{center}
\end{figure}

It is instructive 
to discuss how the subbarrier enhancement of fusion cross sections 
is realized using a schematic model. 
To this end, let us consider a two-channel 
problem for scattering in one-dimension \cite{dasso1983a,dasso1983b} 
and solve the coupled-channels equations in a form of
\begin{equation}
\left[-\frac{\hbar^2}{2m}\frac{d^2}{dx^2}
+
\left(
\begin{array}{cc}
V(x) & F(x) \\
F(x)& V(x)+\epsilon
\end{array}
\right)-E\right]
\left(
\begin{array}{cc}
u_0(x) \\ u_1(x)
\end{array}
\right)=0.
\label{eq:CC}
\end{equation}
Here, $V(x)$ describes a potential barrier and $F(x)$ denotes the coupling 
potential between the two channels. 
$\epsilon$ is the energy of the excited state relative to the 
ground state. 
Assuming that the particle is incident from the right hand side of the potential 
barrier, these equations are solved with the boundary conditions of 
\begin{eqnarray}
u_0(x)&\to& e^{-ik_0x}-R_{0}e^{ik_0x}~~~~~~~(x\to\infty), \\
&\to& T_{0}e^{-ik_0x}~~~~~~~~~~~~~~~~~~~(x\to -\infty),
\end{eqnarray}
and
\begin{eqnarray}
u_1(x)&\to& -\sqrt{\frac{k_0}{k_1}}\,R_{1}e^{ik_1x}~~~~~~~~~~~~~~~~~~(x\to\infty), \\
&\to& \sqrt{\frac{k_0}{k_1}}
\,T_{1}e^{-ik_1x}~~~~~~~~~~~~~~~~~~~(x\to -\infty),
\end{eqnarray}
with $k_0=\sqrt{2mE/\hbar^2}$ and 
$k_1=\sqrt{2m(E-\epsilon)/\hbar^2}$. 
The penetrability of the barrier is then given by 
\begin{equation}
P(E)=|T_0|^2+|T_1|^2. 
\end{equation}
The upper panel of Fig. \ref{fig:cc2ch} shows the penetrability $P(E)$ so obtained with 
a Gaussian barrier given by $V(x)=V_0\,e^{-x^2/2\sigma^2}$. 
The coupling potential is also assumed to have a Gaussian form, 
$F(x)=F_0\,e^{-x^2/2\sigma^2}$. 
The parameters are set to be 
$V_0=100$ MeV, $\sigma=3$ fm, and 
$F_0=3$ MeV, together with 
$\epsilon=1$ MeV and 
$m=29\times 938$ MeV/$c^2$. 
The dashed line shows the result without the coupling (i.e., the case with $F_0=0$),  
while the solid line is obtained by solving 
the coupled-channels equations. 
As in the subbarrier fusion enhancement, one can see that the penetrability is 
enhanced at energies below the barrier. 

When the excitation energy $\epsilon$ is zero, the coupled-channels 
equations (\ref{eq:CC}) can be transformed to two decoupled equations, 
\begin{equation}
\left[-\frac{\hbar^2}{2m}\frac{d^2}{dx^2}
+V(x)\pm F(x)-E\right]u_\pm(x)=0,
\end{equation}
with $u_\pm(x)=(u_0(x)\pm u_1(x))/\sqrt{2}$. 
That is, the wave functions $u_\pm(x)$ are governed by the potentials 
$V(x)\pm F(x)$, one of which lowers the barrier and the other raises the 
barrier. The penetrability is then given by 
\begin{equation}
P(E)=\frac{1}{2}\left[P_0(E;V(x)+F(x))+P_0(E;V(x)-F(x))\right],
\label{eq:cc2ch0}
\end{equation}
where $P_0(E;V(x))$ is the penetrability for a potential barrier 
$V(x)$. Similar to fusion of deformed nuclei, the total penetrability is enhanced 
as compared to the penetrability for the no-coupling case, because of the contribution 
of the lowered barrier is much more significant than the contribution of the higher 
barrier. This remains the same even with a finite excitation energy, 
$\epsilon$ \cite{hagino1997}. 

\section{Fusion Barrier Distributions}

Eq. (\ref{eq:cc2ch0}) can be generalized to cases with more than two barriers as,
\begin{equation}
P(E)=\sum_\alpha w_\alpha P_0(E;V_\alpha(x)),
\label{eq:Pcc}
\end{equation}
where $\alpha$ denotes ``eigen-channels'' with the potential $V_\alpha(x)$ 
and $w_\alpha$ is the weight factor for each eigenchannel. 
That is, the penetrability is given as a weighted sum of the 
penetrability for each eigenchannel $\alpha$. 
In this case, a single barrier is replaced by a set of {\it distributed} barrier. 
The corresponding formula for fusion cross sections reads 
\begin{equation}
\sigma_{\rm fus}(E)=\sum_\alpha w_\alpha \sigma_{\rm fus}^{(0)}(E; V_\alpha(r)). 
\end{equation}
In the case of fusion of deformed nuclei, Eq. (\ref{eq:deffus-cross}), 
the eigen-channel $\alpha$ corresponds to 
the orientation angle $\theta$ 
with the weight factor $w_\theta=2\pi\sin\theta$. 

Since the penetrability $P$ varies from zero to one at energies around 
the barrier height, 
its energy derivative shows a Gaussian-like peak centered at the barrier height 
energy (see the dashed line in the lower panel of Fig. \ref{fig:cc2ch}).  
\footnote{For a classical penetrability, $P(E)=\theta(E-V_b)$, the energy 
derivative is given by a delta function, $dP/dE=\delta(E-V_b)$.} This implies that 
the energy derivative of Eq. (\ref{eq:Pcc}) shows many peaks centered at the 
barrier height for each eigenchannel, and that the height of each peak is proportional 
to the corresponding weight factor, $w_\alpha$. 
This is demonstrated in the lower panel of Fig. \ref{fig:cc2ch} for a 2-channel 
case. 

\begin{figure}[tb]
\begin{center}
\includegraphics*[width=0.7\textwidth]{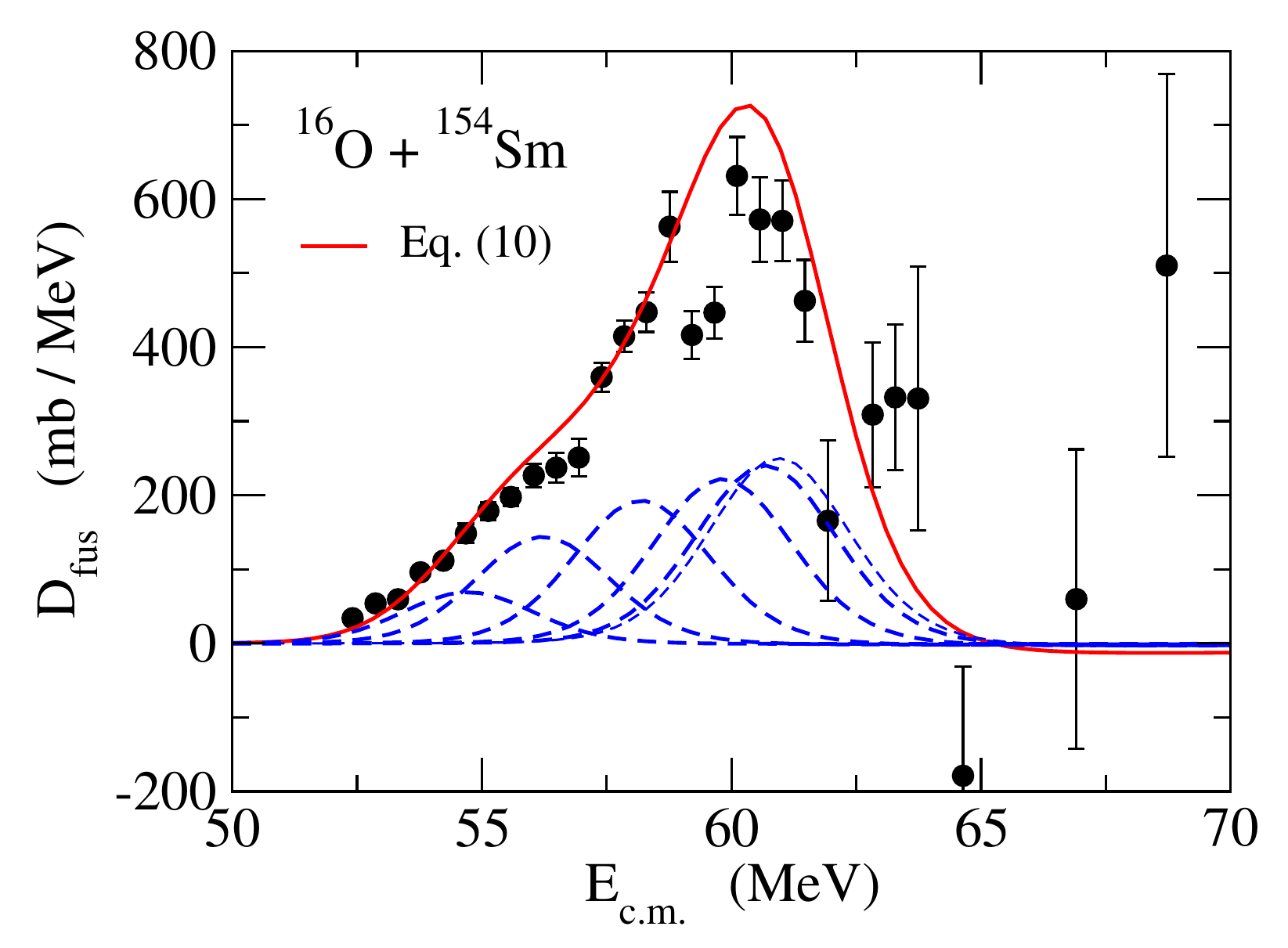}
\caption{
The fusion barrier distribution for the 
$^{16}$O+$^{154}$Sm system. 
The solid line is obtained by averaging all the orientation angles of the 
deformed $^{154}$Sm by Eq. (\ref{eq:deffus-cross}). 
The dashed lines show the contributions of different orientation angles. 
The experimental data are taken from Ref. \cite{Leigh1995}.
}
\label{fig:154sm-bd}
\end{center}
\end{figure}

Noticing the relation given by Eq. (\ref{eq:bd0}), 
one finds that the corresponding quantity for fusion cross sections is 
the second energy derivative of $E\sigma_{\rm fus}(E)$ 
given by 
\begin{equation}
D_{\rm fus}(E)=\frac{d^2(E\sigma_{\rm fus}(E))}{dE^2}.
\end{equation}
This quantity is referred to as the fusion barrier 
distribution \cite{DHRS98,RSS1991}, 
and has been experimentally extracted 
for several systems \cite{DHRS98,Leigh1995}. 
Similar barrier distributions have been extracted using also quasi-elastic scattering 
(that is, a sum of elastic, inelastic, and transfer processes) at backward 
angles \cite{Timmers1995,HR2004}.  
The fusion barrier distribution converts the exponential behavior of fusion 
excitation functions to the linear scale, and is suitable to visualize 
details of the underlying dynamics of subbarrier fusion 
reactions. 
Fig. \ref{fig:154sm-bd} shows the barrier distribution for the 
$^{16}$O+$^{154}$Sm system as an example. 
The solid line shows the barrier distribution obtained with 
Eq. (\ref{eq:deffus-cross}), while the dashed lines show 
the contribution of different orientation angles. 
The fusion barrier distribution is structured because of the distribution 
of many barriers. It has been shown that the shape of the fusion barrier 
distribution is sensitive to the deformation parameters used in 
the calculation \cite{DHRS98,Leigh1995}. 

\begin{figure}[tb]
\begin{center}
\includegraphics*[width=0.45\textwidth]{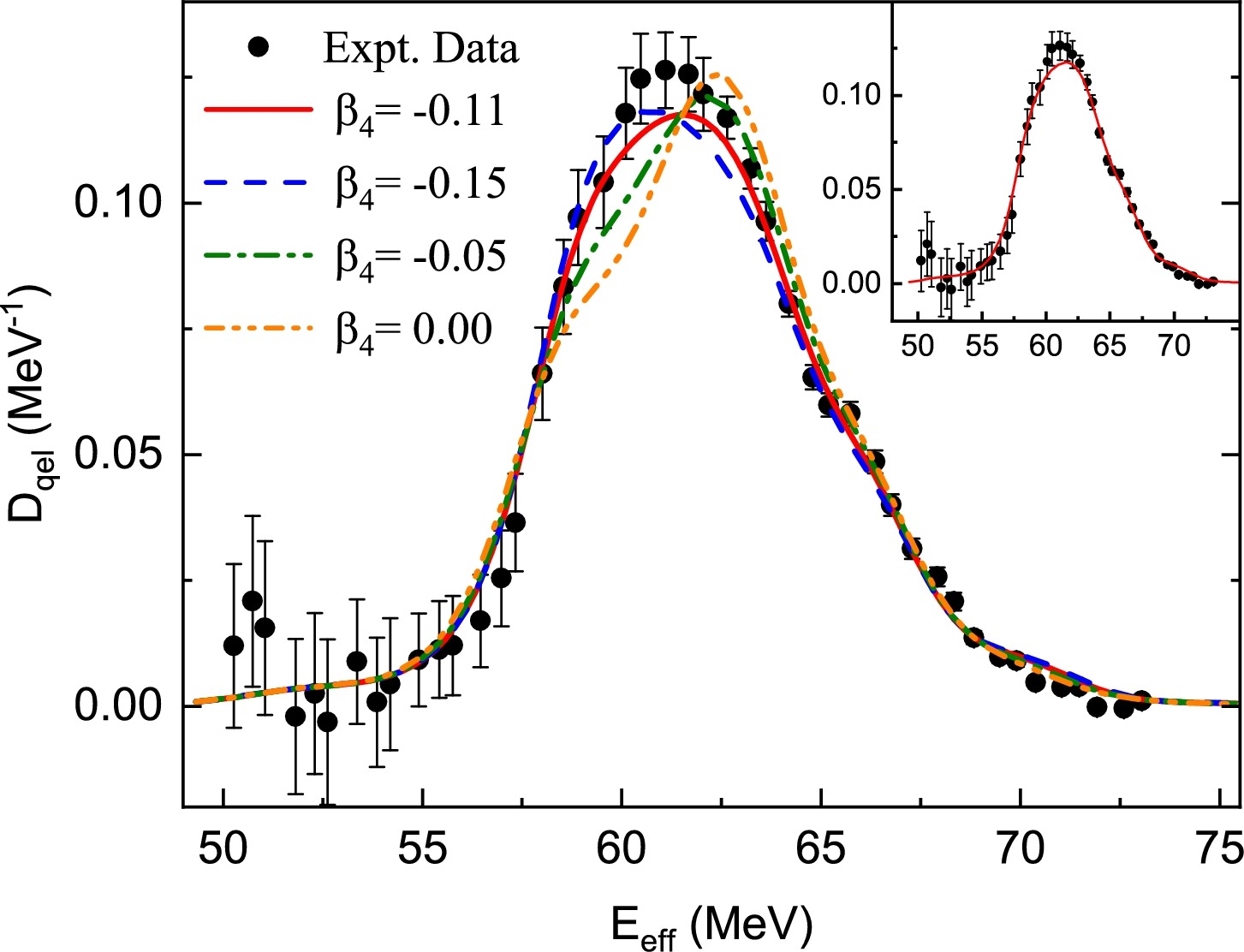}
\includegraphics*[width=0.45\textwidth]{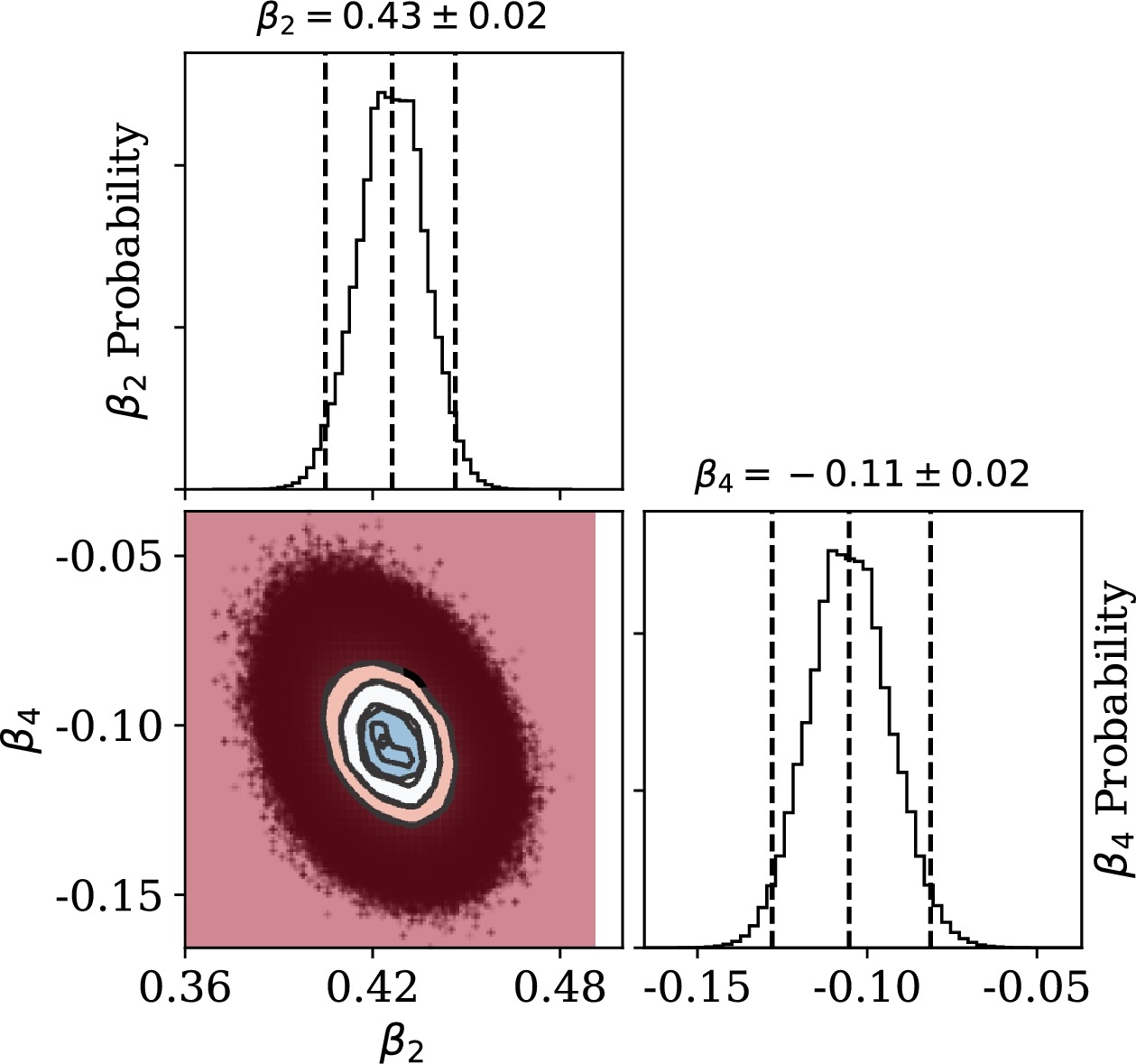}
\caption{The barrier distribution for the $^{24}$Mg+$^{90}$Zr 
system extracted from quasi-elastic scattering at backward angles. 
The right panel shows the posterior probability distribution 
for the Bayesian analysis
for the quadrupole and the hexadecapole deformation parameters of $^{24}$Mg. 
Taken from Ref. \cite{Gupta2020}. 
}
\label{fig:24mg-bd}
\end{center}
\end{figure}

Using such sensitivity of the barrier distribution to the deformation 
parameters, 
the quadrupole and the hexadecapole deformation 
parameters of the 
$^{24}$Mg nucleus have been extracted recently \cite{Gupta2020}. 
To this end, the barrier distribution extracted from the quasi-elastic 
scattering for the $^{24}$Mg+$^{90}$Zr system was analyzed with 
the Bayesian statistics. 
From this analysis, the hexadecapole 
deformation parameter of $^{24}$Mg, $\beta_4=-0.11\pm 0.02$, 
has been determined precisely for the first 
time (see Fig. \ref{fig:24mg-bd}). 

\section{Deep Subbarrier Fusion Hindrance}

At energies well below the Coulomb barrier, that is, at $E\ll V_b$, 
the Wong formula (\ref{eq:wong}) is reduced to 
\begin{equation}
\sigma_{\rm fus}(E)
\sim\frac{\hbar\Omega}{2E}\,R_b^2\,
\exp\left(-\frac{2\pi}{\hbar\Omega}(E-V_b)\right). 
\end{equation}
That is, fusion cross sections fall off exponentially. 
This has been generally observed experimentally. 
However, as the energy decreases further down, 
it has been systematically observed that 
fusion cross sections fall off much steeper  
 \cite{Jiang2002,Back14,jiang2021}. 
This phenomenon has been referred to as {\it deep subbarrier fusion hindrance}. 
It has been considered that the hindrance is attributed to the 
dynamics after two colliding nuclei 
touch with each other \cite{IHI2007b,jiang2021}. 

\begin{figure}[tb]
\begin{center}
\includegraphics*[width=0.7\textwidth]{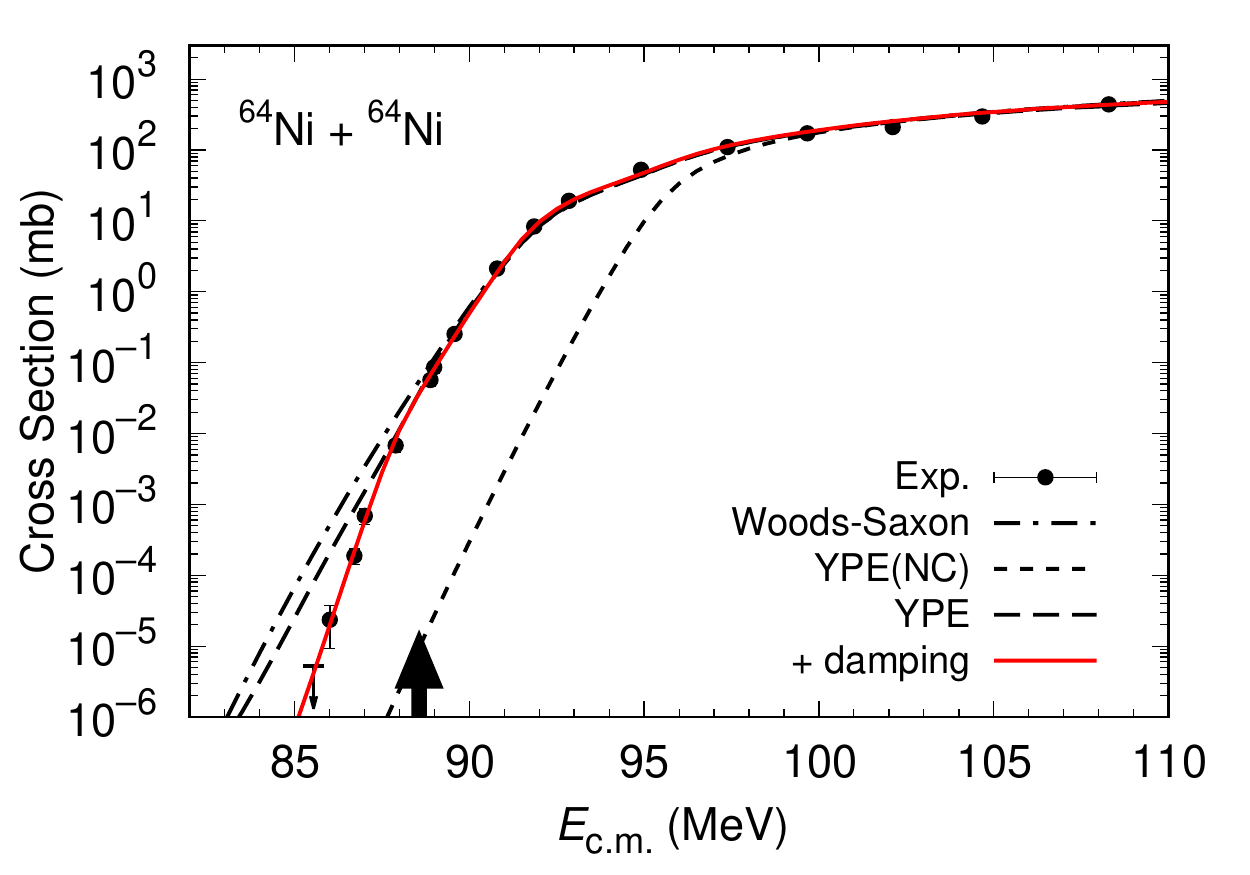}
\caption{The fusion cross sections for the $^{64}$Ni+$^{64}$Ni system.  
The dashed and the dot-dashed lines show the results of 
the standard coupled-channels calculations with two different potentials. 
The solid line is obtained with the adiabatic model for deep 
subbarrier fusion hindrance, which introduces a quenching of coupling strengths 
after the touching. 
The arrow indicates the threshold energy for 
the deep subbarrier fusion hindrance. Taken from Refs. \cite{Ichikawa2015}. 
}
\label{fig:deep-subbarrier}
\end{center}
\end{figure}

As an example, 
Fig. \ref{fig:deep-subbarrier} shows 
the fusion cross sections for the $^{64}$Ni+$^{64}$Ni system. 
The results of the standard coupled-channels calculations are denoted 
by the dashed and the dot-dashed lines. 
These calculations well reproduce the experimental data at energies larger than 
about 89 MeV (see the arrow). 
However, at lower energies the experimental data show hindrance as compared to the 
standard coupled-channels calculation and fall off much steeper. 
The solid line models the deep subbarrier hindrance by quenching the coupling 
strengths after two nuclei touch each other \cite{Ichikawa2015} (see Ref. \cite{ME2006} 
for another modelling of deep subbarier fusion hindrance, which introduces a repulsive 
core to an internucleus potential). 
This calculation well accounts for the data, clearly indicating an importance of the 
dynamics of a transition from a two-body system with two separate nuclei 
to a one-body mono-nuclear 
system after the touching. See Ref. \cite{jiang2021} for a recent review article 
on this topic. 

\section{Fusion of Neutron-Rich Nuclei} 

One of the main research fields in modern nuclear physics is physics of 
unstable nuclei, especially neutron-rich nuclei far from the stability line. 
Those nuclei are weakly bound and are characterized by a spatially extended 
density distribution. 
It is likely that excited states of those nuclei are in the continuum 
spectrum and thus the breakup process plays an important role when 
such nuclei are used either as a projectile or as a target in nuclear reactions. 

\begin{figure}[tb]
\begin{center}
\includegraphics*[width=1\textwidth]{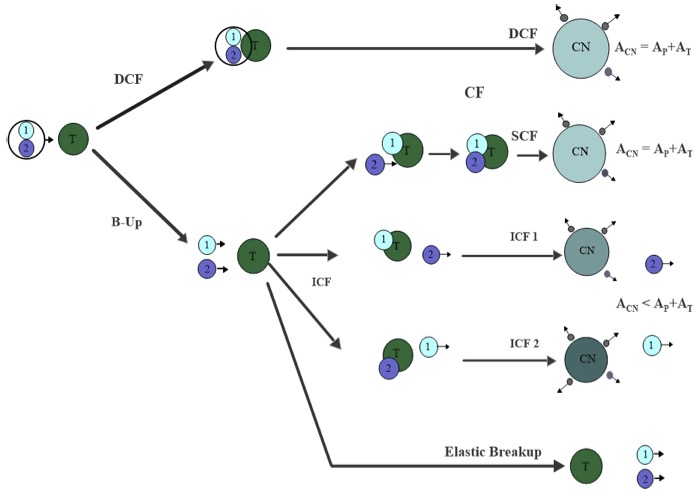}
\caption{
A schematic illustration of fusion dynamics in the presence of breakup of the 
projectile. CF and ICF refer to the complete fusion and the incomplete fusion, 
respectively. The complete fusion is further subdivided into the direct 
complete fusion (DCF) and the sequential complete fusion (SCF) processes. 
Taken from Ref. \cite{canto2006}. 
}
\label{fig:breakup-fusion}
\end{center}
\end{figure}

In fusion of weakly-bound nuclei,  
several effects may interplay with each other. 
Those are:
\begin{enumerate} 

\item 
a lowering of the 
Coulomb barrier 
due to the extended density distribution 
\cite{TS1991}, 

\item the breakup process, which may hinder fusion cross sections since the 
lowering of the Coulomb barrier disappears. At the same time, it may also 
enhance fusion cross 
sections if couplings to a breakup channel 
dynamically lowers the Coulomb barrier \cite{DV1994,HVD00,DTT2002}, 

\item 
the transfer processes. Since 
the $Q$-value is positive for neutron-rich nuclei, 
it may significantly affect the fusion process \cite{CCS18,moschini2021}. 
\end{enumerate} 
Furthermore, the breakup process significantly complicates the reaction dynamics of 
complete fusion and incomplete fusion in a non-trivial 
way (see Fig. \ref{fig:breakup-fusion}). 
Here, the complete fusion is the process in which all the breakup fragments 
are absorbed by a target nucleus, while the incomplete fusion refers to the process 
in which only a part of the breakup fragments is absorbed. 
A theoretical model which coherently 
incorporates all of these effects has still yet to be 
developed, even though the continuum discretized coupled-channels (CDCC) method 
has been developed for the breakup process \cite{yahiro2012} (notice also that there have been 
recent developments in theoretical descriptions of 
inclusive breakup processes \cite{lei2019a,lei2019b,rangel2020,cortes2020}). 
See Refs. \cite{hagino2021,canto2006,canto2015} for review articles on 
fusion of weakly bound nuclei. 
It is worth noticing that 
fusion of neutron-rich nuclei is important for nuclear 
astrophysics \cite{CH2008,Steiner2012} as well as 
for superheavy elements \cite{Loveland2007}. 

\section{Fusion Reactions for Superheavy Nuclei}

\subsection{Superheavy Nuclei}

The elements heavier than plutonium (the atomic number 
$Z=94$) are all unstable and do not 
exist in nature. 
Yet, one can artificially synthesize them using nuclear 
reactions. There have been continuous efforts since the 1950s, 
and the elements up to $Z=118$ have been 
synthesized by now. 
The transactinide elements, that is, the elements with $Z\geq 104$, are 
referred to as superheavy elements and  
have 
attracted lots of attention in recent years 
\cite{HM2000,HHO2013,DHNO2015,HDS2021,Nazarewicz2018,Giuliani2019,H2019}. 

One of the main motivations to study superheavy elements, in addition to 
synthesizing new elements, is to explore the island of stability, which was 
theoretically predicted some 50 years ago \cite{myers1966,sobiczewski1966}. 
While heavy nuclei in the transactinide region are unstable against 
alpha decay and spontaneous fission, the shell effect due to magic 
numbers can stabilize a certain number of nuclei in that region. 
The predicted proton and neutron magic numbers are $Z=114$ and 
$N=184$, respectively \cite{myers1966,sobiczewski1966}. 
The region around these magic numbers is 
referred to as {\it the island of stability}, where nuclei may have a life time 
as long as 10$^3$ years \cite{koura2018}.
More modern Hartree-Fock calculations have also predicted 
$(Z,N)=(114,184), (120,172)$, and (126,184) 
for candidates for the next double magic nucleus 
beyond $^{208}$Pb \cite{bender1999}. 

The island of stability has not yet been reached experimentally. 
In fact, the heaviest Fl element ($Z=114$) synthesized so far 
is $^{289}_{175}$Fl \cite{oganessian04-2}, 
which is 9 neutrons less from the 
the predicted 
magic number, $N=184$. 
This implies that neutron-rich beams are indispensable in order to 
reach the island of stability. 
An experimental technique has yet to be developed 
to deal with low intensity 
of such beams. 

\subsection{Heavy-Ion Fusion Reactions for Superheavy Nuclei}

\begin{figure}[tb]
\begin{center}
\includegraphics*[width=0.7\textwidth]{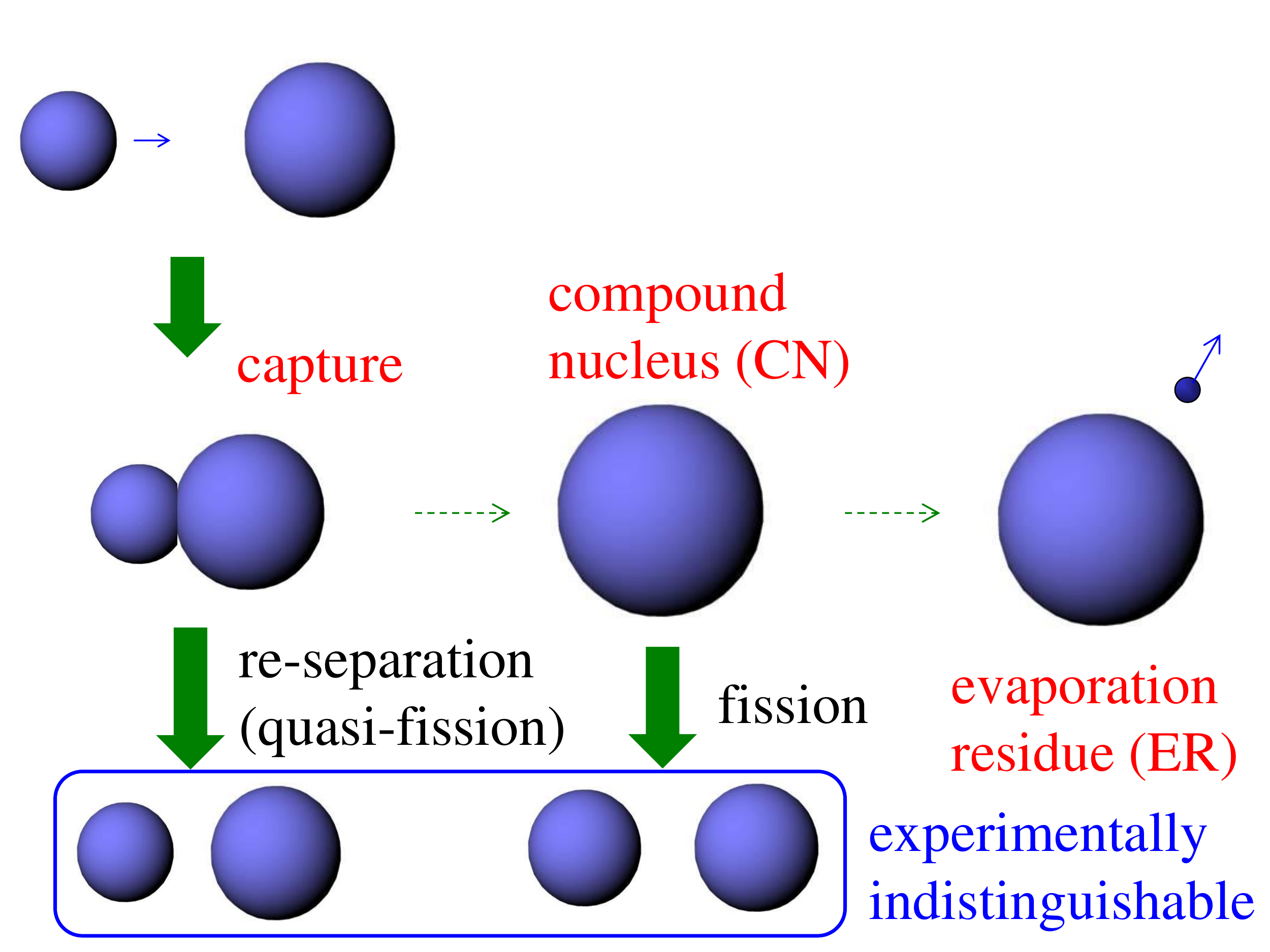}
\caption{
A schematic illustration of heavy-ion fusion reactions to synthesize 
superheavy nuclei.}
\label{fig:she-overview}
\end{center}
\end{figure}

Heavy-ion fusion reactions at energies around the 
Coulomb barrier have been used as 
a standard tool 
to synthesize those superheavy elements \cite{HM2000,HHO2013}. 
Figure \ref{fig:she-overview} schematically illustrates fusion reactions 
to form superheavy nuclei (see also Fig. \ref{fig:overview}). 
In the first phase of reaction, two nuclei approach to each other to reach 
the touching 
configuration after the Coulomb barrier is overcome. 
A compound nucleus is 
formed almost automatically for medium-heavy systems once the touching configuration 
is achieved. In contrast, 
in the superheavy region, there is a 
huge probability for the touching configuration to reseparate due to a 
strong Coulomb repulsion between the two nuclei. 
This process is referred to as {\it quasi-fission}. 
Furthermore, even if a compound nucleus is formed with a small probability, 
it decays most likely by fission, again due to the strong 
Coulomb interaction. 
For heavy systems, 
quasi-fission 
characteristics significantly overlap with fission of the compound nucleus, 
and a detection of fission events itself does not guarantee a formation of the 
compound nucleus. 
Therefore, a formation of superheavy elements has been identified by measuring 
evaporation residues. 
These are extremely rare events, in which a compound nucleus is 
survived against fission. 

\begin{figure}[tb]
\begin{center}
\includegraphics*[width=0.5\textwidth]{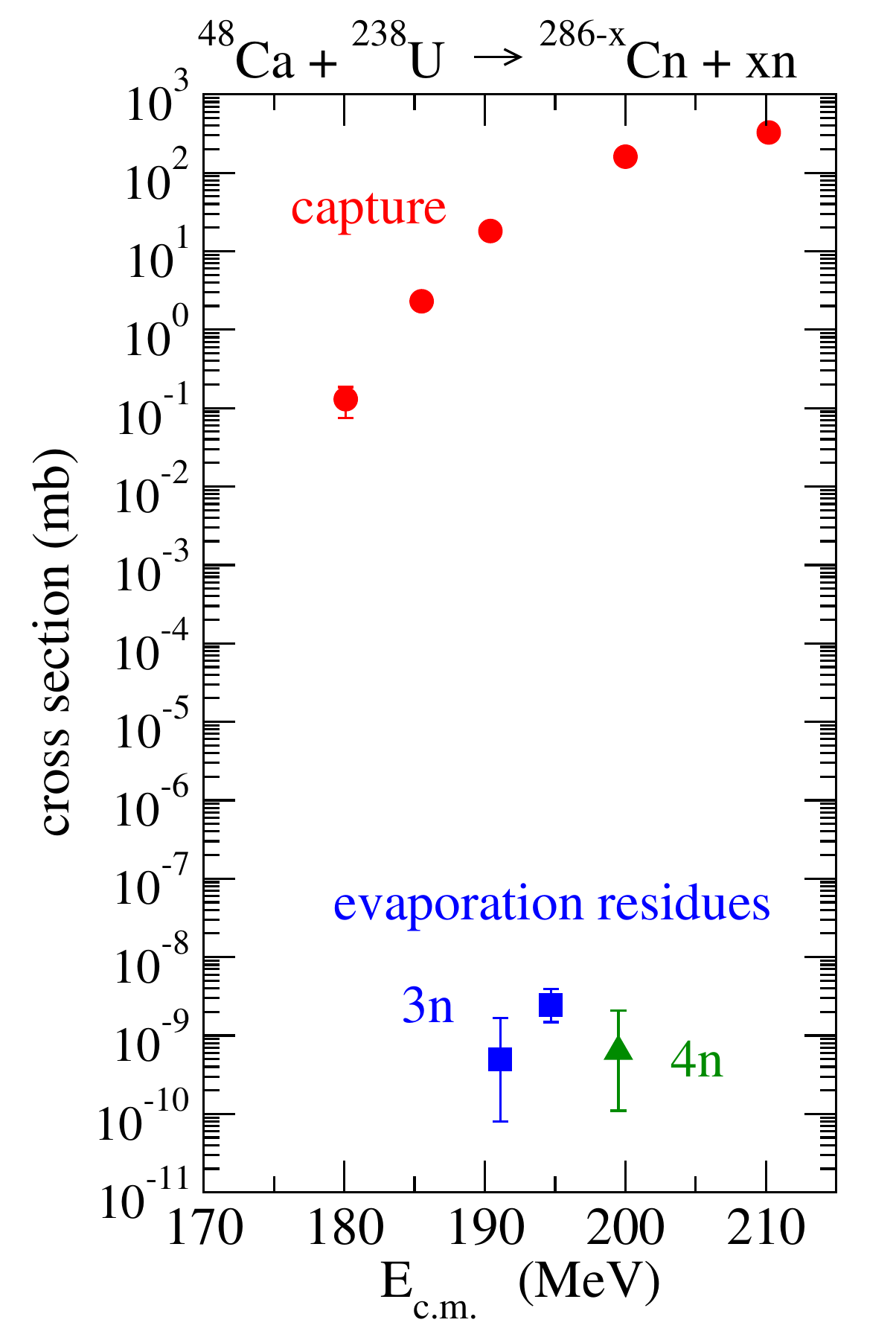}
\caption{The experimental evaporation residue cross sections 
for the $^{48}$Ca+$^{238}$U reaction leading to the formation of Cn ($Z=112$) element. 
The filled circles denote the capture cross sections \cite{kozulin2014} 
to form the touching 
configuration. The filled squares and triangles show 
the evaporation residue cross sections \cite{oganessian2000,oganessian2004}, 
for which the former and the latter 
correspond to the 3$n$ (emission of 3 neutrons) and the 
4$n$ (emission of 4 neutrons) channels, respectively. }
\label{fig:48ca238u}
\end{center}
\end{figure}

As an example, Fig. \ref{fig:48ca238u}  
shows the measured cross sections for 
the $^{48}$Ca+$^{238}$U reaction 
forming the Cn ($Z=112$) element. The filled circles show the capture 
cross sections \cite{kozulin2014} 
to form the touching configuration shown in Fig. \ref{fig:she-overview}. 
On the other hand, the filled squares and triangles denote the evaporation 
residue cross sections for emissions of 3 and 4 neutrons, respectively 
\cite{oganessian2000,oganessian2004}. 
One can observe that the evaporation residue cross sections 
are indeed much smaller than the capture cross sections, 
by about 11 orders of magnitude. 

\subsection{Theoretical modelings}

Based on the time-scale of 
each process, the formation process of evaporation residues  
can be conceptually divided into a sequence of the following 
three processes (see Fig. \ref{fig:she-overview}). 
The first phase is a process in which two separate nuclei form the touching 
configuration after overcoming the Coulomb barrier. 
After two nuclei touch with each other, a huge number of nuclear intrinsic 
motions are activated 
and the energy for the relative motion of the colliding nuclei 
is quickly dissipated to internal energies. 
Because of the strong Coulomb interaction, 
the touching configuration appears outside a fission 
barrier, which has to be 
thermally activated to form a 
a compound nucleus against a severe competition to the quasi-fission process. 
The Langevin approach has often been used to describe 
this process \cite{Swiatecki2005,ABGW2000,SKA2002,Swiatecki2003,AO2004,ZG2015}. 
The third process is a statistical decay of the compound 
nucleus \cite{KEWPIE2}, with strong competitions  
between fission and particle emissions (i.e., evaporations). 
Here, the fission barrier height is one of the most important parameters 
which significantly affect evaporation residue cross sections \cite{Boilley16}. 

For a given partial wave $l$, the probability for a formation of 
an evaporation residue is given as a product of the probability for each 
of the three processes, $P_l$, $P_{\rm CN}$, and $W_{\rm sur}$, that is, 
\begin{equation}
P_{\rm ER}(E,l)=P_l(E)P_{\rm CN}(E,l)W_{\rm sur}(E^*,l),
\end{equation}
where $E$ and $E^*$ are the bombarding energy in the center of mass 
frame and the excitation energy of the compound nucleus, respectively. 
Cross sections for a formation of evaporation residues are then given by 
\begin{equation}
\sigma_{\rm ER}(E)=\frac{\pi}{k^2}\sum_l (2l+1)
P_l(E)P_{\rm CN}(E,l)W_{\rm suv}(E^*,l).
\label{eq:sigma_ER}
\end{equation}

For medium-heavy systems, the probability for the second phase, 
$P_{\rm CN}$, is almost unity, and Eq. (\ref{eq:sigma_ER}) is reduced to 
Eq. (\ref{eq:crosssections0}) when fission cross sections are added to it. 
In contrast, for superheavy nuclei, 
$P_{\rm CN}$ is significantly smaller than unity. 
As has been mentioned, this quantity cannot be determined experimentally, 
which causes 
large ambiguities in 
theoretical calculations. 

\subsection{Hot versus Cold Fusion Reactions}

\begin{figure}[bt]
\begin{center}
\includegraphics[width=0.7\linewidth]{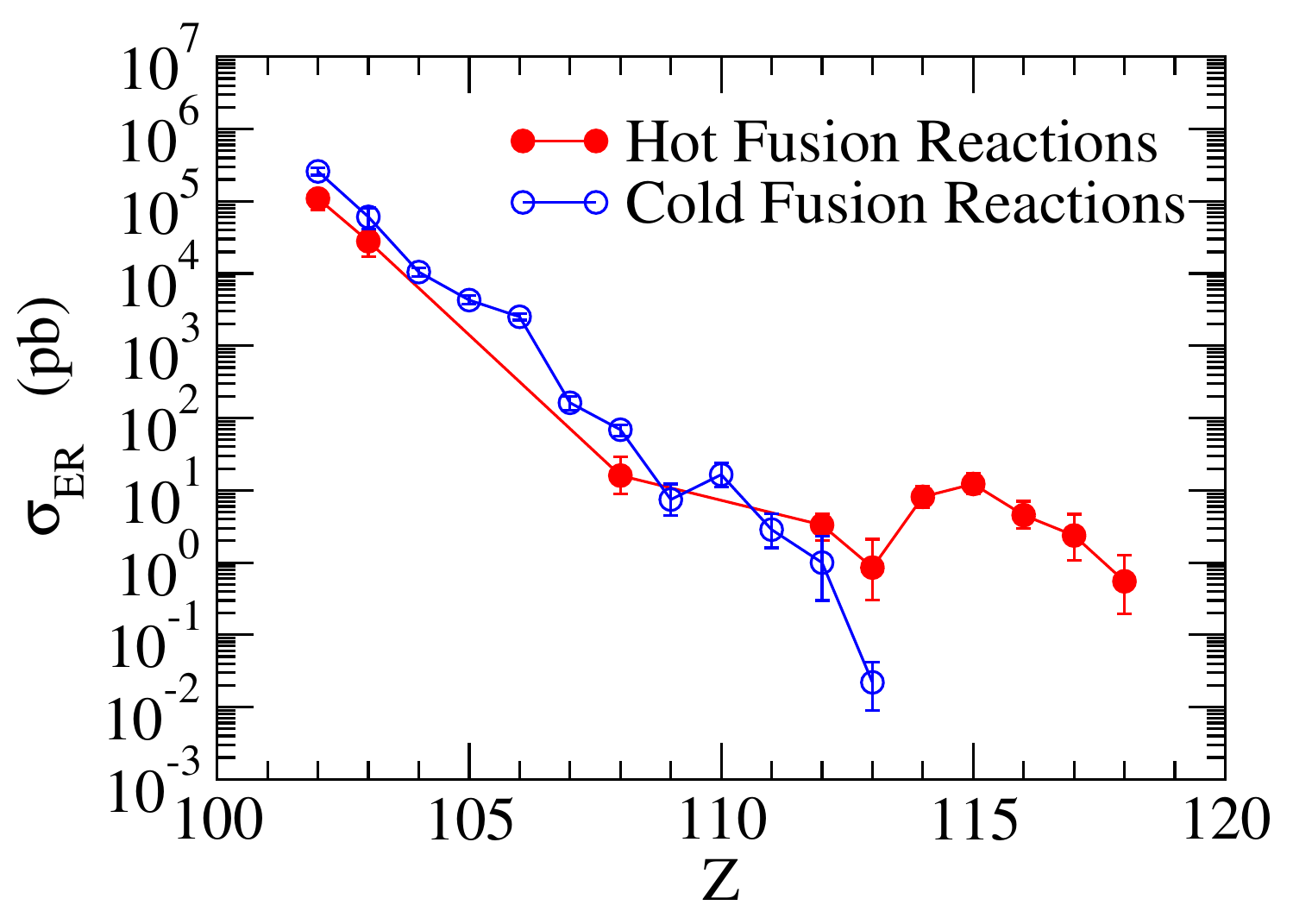}
\end{center}
\caption{The measured evaporation residue cross sections 
as a function of the atomic number $Z$ of a compound nucleus. 
The filled circles denote the results of the hot fusion reactions with 
$^{48}$Ca projectile. 
The open circles show the 
fusion cross sections obtained with the cold fusion reactions, 
in which $^{208}$Pb or $^{209}$Bi 
nuclei 
are used as a target.  
The maximum of a sum of the 3$n$ and 4$n$ 
cross sections 
and that of the 1$n$ cross sections 
are shown for each $Z$ for the hot and the cold fusion reactions, respectively. 
The experimental data 
are taken from Refs. \cite{NRV,oganessian2013,morita2004,morita2007,morita2012}.}
\label{fig:hot-cold-fusion}
\end{figure}

Since a formation probability of evaporation residues is extremely small, 
it is important to choose appropriate combinations of the projectile and 
the target nuclei in order to efficiently synthesize superheavy elements. 
For this purpose, two different experimental strategies have been employed.  
In the so called {\it cold fusion reactions}, 
$^{208}$Pb and $^{209}$Bi are used for the target nuclei so that 
the compound nucleus is formed 
with relatively low excitation energies.  
The competition 
between neutron emissions and fission can be 
minimized, 
which in turn maximizes $W_{\rm sur}$ in Eq. (\ref{eq:sigma_ER}) \cite{HM2000,HHO2013}. 
An advantage of this strategy is that 
alpha decays of the evaporation residues end up in the known region of 
nuclear chart, and thus superheavy elements can be identified unambiguously. 
On the other hand, in the so called {\it hot fusion reactions}, 
the neutron-rich double magic nucleus $^{48}$Ca has 
been used as a projectile 
to optimize the formation probability of the compound nucleus, 
$P_{\rm CN}$ \cite{HM2000,HHO2013}. 
This strategy has been successfully employed by the experimental group at 
Dubna, led by Oganessian, to synthesize superheavy 
elements up to $Z=118$. 

Figure \ref{fig:hot-cold-fusion}
shows the measured evaporation residue cross sections 
for the hot fusion reactions (the filled circles) and for the cold fusion 
reactions (the open circles). 
For the cold fusion reactions, the cross sections drop rapidly as a 
function of $Z$ of the compound nucleus. It would therefore be difficult 
to go beyond Nihonium using this strategy. In contrast, for the hot fusion 
reactions, the cross sections remain relatively large between $Z=113$ and 118. 
This may be due to the fact that 
the survival probability, $W_{\rm sur}$, is increased because 
the compound nuclei formed are in the 
proximity of the predicted island of 
stability \cite{myers1966,sobiczewski1966}. 
An increase of nuclear dissipation at 
high temperatures may also play a role \cite{yanez2014}. 

\subsection{Role of deformation in hot fusion reactions}

The incident energy for fusion reactions to synthesize superheavy nuclei 
is usually taken at energies slightly above the Coulomb barrier. 
This is because the compound nucleus formed has to be as cold as possible, 
but yet the capture probability, $P_l(E)$, has to be large enough. 

In the hot fusion reactions, 
with the $^{48}$Ca projectile, 
the corresponding target nuclei are in the actinide region, in which 
the nuclei are well deformed in the ground state. 
It has been argued that 
the collision with $\theta=\pi/2$ (i.e., the ``side collision'', 
see Fig. \ref{fig:defpot2}) plays an important role in the hot fusion, 
since the touching configuration is more compact 
than that formed with the ``tip collision'' with $\theta=0$,  
and thus the effective barrier height 
for the diffusion process is low \cite{Hinde1995,hagino2018}. 

\begin{figure}[tb]
\begin{center}
\includegraphics*[width=0.7\textwidth]{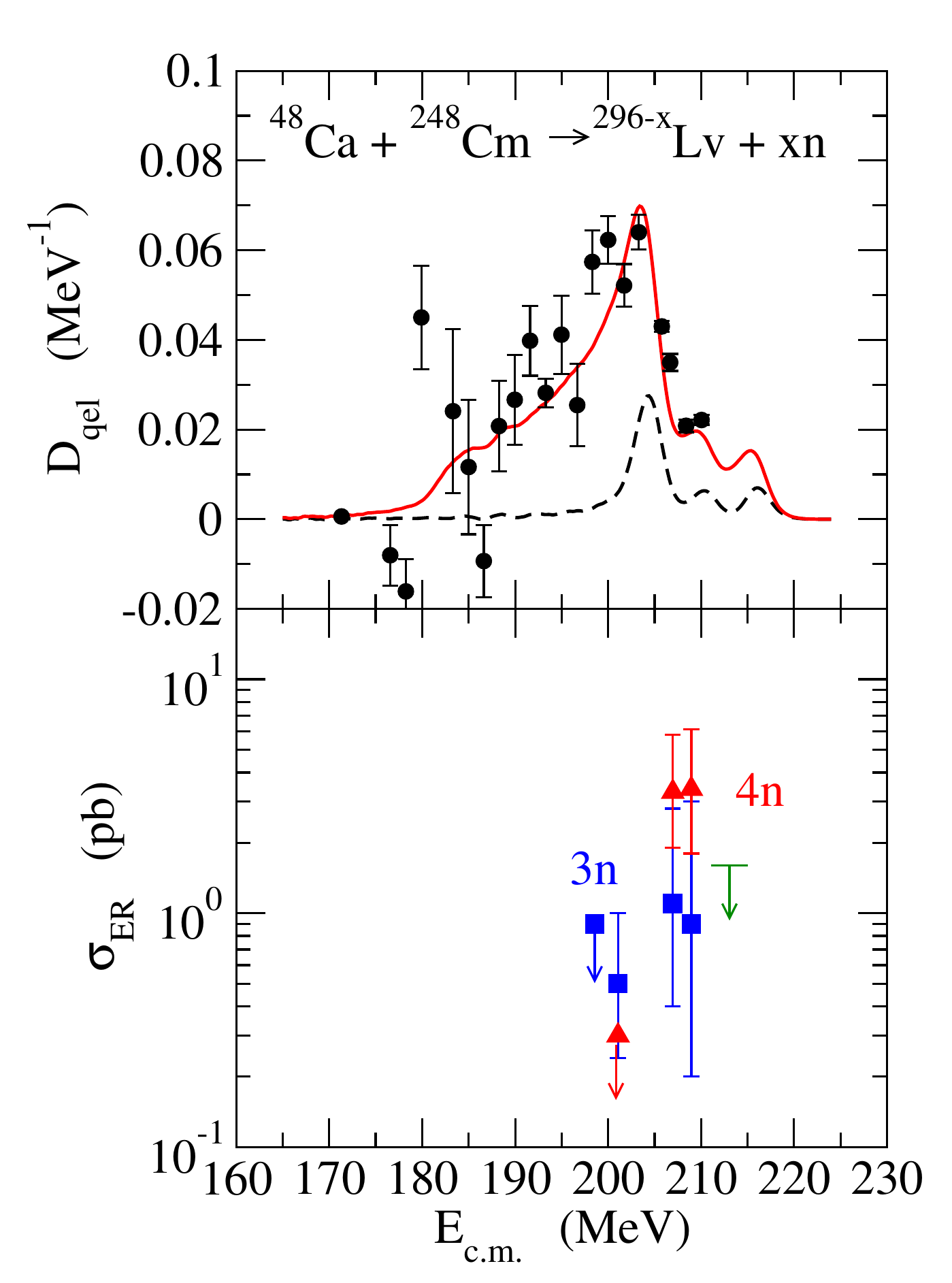}
\caption{(Upper panel) The barrier distribution for the 
capture process for the $^{48}$Ca+$^{248}$Cm system extracted from quasi-elastic 
scattering at backward angles. 
The solid line is obtained with the coupled-channels calculation which 
takes into account the deformation of the $^{248}$Cm target. 
The octupole phonon 
excitation of $^{48}$Ca and a one-neutron transfer process are also taken 
into account. 
The dashed line shows the contribution of the side collision with $\theta=\pi/2$. 
The experimental data are taken from Ref. \cite{Tanaka2018}. 
(Lower panel) 
The evaporation residue cross sections for the same system. 
The experimental data are taken from Ref. \cite{oganessian2004,Hofmann2012}. 
}
\label{fig:48ca238cm}
\end{center}
\end{figure}

Recently, barrier distributions for the capture process 
have been extracted for several 
hot fusion systems and the notion of compactness has been confirmed 
experimentally for the first time \cite{Tanaka2018,Tanaka2020}. 
As an example, the top panel of Fig. \ref{fig:48ca238cm} shows 
the experimental barrier distribution 
for the $^{48}$Ca+$^{248}$Cm system \cite{Tanaka2018} and its comparison to the 
coupled-channels calculations which take into account 
the deformation of the $^{248}$Cm nucleus. 
The coupled-channels calculation provides information on the energy 
region corresponding to the side collision, 
which is indicated by 
the dashed line in the figure. The measured evaporation residue cross sections 
$\sigma_{\rm ER}$  
for this system \cite{oganessian2004,Hofmann2012} are shown in the lower panel of the figure. 
One can clearly see that the peak of evaporation residue cross sections, 
$\sigma_{\rm ER}$, appears in the energy 
region corresponding to the side collision, which is compatible with the notion of 
compactness.

\bigskip

\noindent
{\bf Acknowledgements} 

I thank L. Felipe Canto for his 
careful reading of the mansucrpt and useful suggestions. 
I also thank Tomoya Naito for his help in preparing Fig. 6. 
This work was 
supported in part by JSPS KAKENHI
Grant Numbers JP19K03861 and JP21H00120.

\bibliography{handbook-hagino}

\begin{thebibliography}{10}

\bibitem{bohr1936}
N.~Bohr.
\newblock {\em Nature}, 137:351, 1936.

\bibitem{asghar1966}
M.~{Asghar et al.}
\newblock {\em Nucl. Phys.}, 85:305, 1966.

\bibitem{bender2020}
M.~{Bender et al.}
\newblock {\em J. of Phys. G}, 47:113002, 2002.

\bibitem{morita2004}
K.~{Morita et al.}
\newblock {\em J. of Phys. Soc. Jpn.}, 73:2593, 2004.

\bibitem{morita2007}
K.~{Morita et al.}
\newblock {\em J. of Phys. Soc. Jpn.}, 76:5001, 2007.

\bibitem{morita2012}
K.~{Morita et al.}
\newblock {\em J. of Phys. Soc. Jpn.}, 81:3201, 2012.

\bibitem{bass1980}
R.~Bass.
\newblock {\em Nucler Reactions with Heavy-Ions}.
\newblock Springer-Verlag, Berlin, 1980.

\bibitem{BT98}
A.~B. Balantekin and N.~Takigawa.
\newblock {\em Rev. Mod. Phys.}, 70:77, 1998.

\bibitem{hagino2012}
K.~Hagino and N.~Takigawa.
\newblock {\em Prog. Theo. Phys.}, 128:1061, 2012.

\bibitem{DHRS98}
M.~Dasgupta, D.~J. Hinde, N.~Rowley, and A.M. Stefanini.
\newblock {\em Annu. Rev. Nucl. Part. Sci.}, 48:401, 1998.

\bibitem{Back14}
B.~B. Back, H.~Esbensen, C.~L. Jiang, and K.~E. Rehm.
\newblock {\em Rev. Mod. Phys.}, 86:317, 2014.

\bibitem{MS17}
G.~Montagnoli and A.~M. Stefanini.
\newblock {\em Eur. Phys. J. A}, 53:169, 2017.

\bibitem{jiang2021}
C.~L. Jiang, B.~B. Back, K.~E. Rehm, K.~Hagino, G.~Montagnoli, and A.~M.
  Stefanini.
\newblock {\em Eur. Phys. J. A}, 57:235, 2021.

\bibitem{frobrich1996}
P.~Fr\"obrich and R.~Lipperheide.
\newblock {\em Theory of Nuclear Reactions}.
\newblock Clarendn Press, Oxford, 1996.

\bibitem{bertulani2004}
C.~A. Bertulani and P.~Danielewicz.
\newblock {\em Introduction to Nuclear Reactions}.
\newblock IOP Publishing, Bristol, UK, 2004.

\bibitem{thompson2009}
I.~J. Thompson and F.~M. Nunes.
\newblock {\em Nuclear Reactions for Astrophysics}.
\newblock Cambridge University Press, Cambridge, 2009.

\bibitem{canto2013}
L.~F. Canto and M.~S. Hussein.
\newblock {\em Scattering Theory of Molecules, Atoms and Nuclei}.
\newblock World Scientific Publishing Co. Pte. Ltd., Singapore, 2013.

\bibitem{Wong1973}
C.~Y. Wong.
\newblock {\em Phys. Rev. Lett.}, 31:766, 1973.

\bibitem{RH2015}
N.~Rowley and K.~Hagino.
\newblock {\em Phys. Rev. C}, 91:044617, 2015.

\bibitem{beckerman1985}
M.~Beckerman.
\newblock {\em Phys. Rep.}, 129:145, 1985.

\bibitem{beckerman1988}
M.~Beckerman.
\newblock {\em Rep. Prog. Phys.}, 51:1047, 1988.

\bibitem{switkowski1977}
Z.~E. Switkowski, R.~G. Stokstad, and R.~M. Wieland.
\newblock {\em Nucl. Phys. A}, 279:502, 1977.

\bibitem{Leigh1995}
J.~R. Leigh, M.~Dasgupta, D.~J. Hinde, J.~C. Mein, C.~R. Morton, R.~C. Lemmon,
  J.~P. Lestone, J.~O. Newton, H.~Timmers, J.~X. Wei, and N.~Rowley.
\newblock {\em Phys. Rev. C}, 52:3151, 1995.

\bibitem{Naito2021}
T.~Naito, S.~Endo, K.~Hagino, and Y.~Tanimura.
\newblock {\em J. of Phys. B}, 54:165201, 2021.

\bibitem{leigh1993}
J.~R. Leigh, N.~Rowley, R.~C. Lemmon, D.~J. Hinde, J.~O. Newton, J.~X. Wei,
  J.~C. Mein, C.~R. Morton, S.~Kuyucak, and A.~T. Kruppa.
\newblock {\em Phys. Rev. C}, 47:R437, 1993.

\bibitem{hagino2021}
K.~Hagino, K.~Ogata, and A.~M. Moro.
\newblock {\em to be published}, 2021.

\bibitem{tamura1965}
T.~Tamura.
\newblock {\em Rev. Mod. Phys.}, 37:679, 1965.

\bibitem{satchler1983}
G.~R. Satchler.
\newblock {\em Direct Nuclear Reactions}.
\newblock Clarendn Press, Oxford, 1983.

\bibitem{broglia2004}
R.~A. Broglia and A.~Winther.
\newblock {\em Heavy-ion Reactions}.
\newblock Westview Press, Cambridge, MA, 2004.

\bibitem{beckerman1981}
M.~Beckerman, J.~Ball, H.~Enge, M.~Salomaa, A.~Sperduto, S.~Gazes, A.~DiRienzo,
  and J.~D. Molitoris.
\newblock {\em Phys. Rev. C}, 23:1581, 1981.

\bibitem{ecis}
J.~Raynal.
\newblock {\em Saclay Report No. DPh-T 69/42 (unpublished)}.

\bibitem{lepine-szily2021}
A.~L\'epine-Szily and R.~Lichtenth\"aler.
\newblock {\em Euro. Phys. J. A}, 57:99, 2021.

\bibitem{fresco}
I.~J. Thompson.
\newblock {\em Comput. Phys. Rep.}, 7:167, 1988.

\bibitem{ccfull}
K.~Hagino, N.~Rowley, and A.~T. Kruppa.
\newblock {\em Comput. Phys. Comm.}, 123:143, 1999.

\bibitem{dasso1983a}
C.~H. Dasso, S.~Landowne, and A.~Winther.
\newblock {\em Nucl. Phys. A}, 405:381, 1983.

\bibitem{dasso1983b}
C.~H. Dasso, S.~Landowne, and A.~Winther.
\newblock {\em Nucl. Phys. A}, 407:221, 1983.

\bibitem{hagino1997}
K.~Hagino, N.~Takigawa, and A.~B. Balantekin.
\newblock {\em Phys. Rev. C}, 56:2104, 1997.

\bibitem{RSS1991}
N.~Rowley, G.~R. Satchler, and P.~H. Stelson.
\newblock {\em Phys. Lett. B}, 254:25, 1991.

\bibitem{Timmers1995}
H.~Timmers, J.R. Leigh, M.~Dasgupta, D.J. Hinde, R.C. Lemmon, J.C. Mein, C.R.
  Morton, J.O. Newton, and N.~Rowley.
\newblock {\em Nucl. Phys. A}, 584:190, 1995.

\bibitem{HR2004}
K.~Hagino and N.~Rowley.
\newblock {\em Phys. Rev. C}, 69:054610, 2004.

\bibitem{Gupta2020}
Y.K. Gupta, B.K. Nayak, U.~Garg, K.~Hagino, K.B. Howard, N.~Sensharma,
  M.~Şenyiğit, W.P. Tan, P.D. O'Malley, M.~Smith, Ramandeep Gandhi,
  T.~Anderson, R.J. deBoer, B.~Frentz, A.~Gyurjinyan, O.~Hall, M.R. Hall,
  J.~Hu, E.~Lamere, Q.~Liu, A.~Long, W.~Lu, S.~Lyons, K.~Ostdiek, C.~Seymour,
  M.~Skulski, and B.~{Vande Kolk}.
\newblock {\em Phys. Lett. B}, 806:135473, 2020.

\bibitem{Jiang2002}
C.~L. Jiang, H.~Esbensen, K.~E. Rehm, B.~B. Back, R.~V.~F. Janssens, J.~A.
  Caggiano, P.~Collon, J.~Greene, A.~M. Heinz, D.~J. Henderson, I.~Nishinaka,
  T.~O. Pennington, and D.~Seweryniak.
\newblock {\em Phys. Rev. Lett.}, 89:052701, 2002.

\bibitem{IHI2007b}
T.~Ichikawa, K.~Hagino, and A.~Iwamoto.
\newblock {\em Phys. Rev. C}, 75:064612, 2007.

\bibitem{Ichikawa2015}
T.~Ichikawa.
\newblock {\em Phys. Rev. C}, 92:064604, 2015.

\bibitem{ME2006}
\ifmmode \mbox{\c{S}}\else \c{S}\fi{}. Mi\ifmmode~\mbox{\c{s}}\else
  \c{s}\fi{}icu and H.~Esbensen.
\newblock {\em Phys. Rev. Lett.}, 96:112701, 2006.

\bibitem{canto2006}
L.~F. Canto, P.~R.~S. Gomes, R.~Donangelo, and M.~S. Hussein.
\newblock {\em Phys. Rep.}, 424:1, 2006.

\bibitem{TS1991}
N.~Takigawa and H.~Sagawa.
\newblock {\em Phys. Lett. B}, 265:23, 1991.

\bibitem{DV1994}
C.~H. Dasso and A.~Vitturi.
\newblock {\em Phys. Rev. C}, 50:R12, 1994.

\bibitem{HVD00}
K.~Hagino, A.~Vitturi, C.~H. Dasso, and S.~M. Lenzi.
\newblock {\em Phys. Rev. C}, 61:037602, 2000.

\bibitem{DTT2002}
A.~Diaz-Torres and I.~J. Thompson.
\newblock {\em Phys. Rev. C}, 65:024606, 2002.

\bibitem{CCS18}
K.~S. Choi, M.~K. Cheoun, W.~So, K.~Hagino, and K.~Kim.
\newblock {\em Phys. Lett. B}, 780:455, 2018.

\bibitem{moschini2021}
L.~Moschini, A.~M. Moro, and A.~Vitturi.
\newblock {\em Phys. Rev. C}, 103:014604, 2021.

\bibitem{yahiro2012}
M.~Yahiro, K.~Ogata, T.~Matsumoto, and K.~Minomo.
\newblock {\em Prog. Theo. Exp. Phys.}, 2012:01A206, 2012.

\bibitem{lei2019a}
J.~Lei and A.~M. Moro.
\newblock {\em Phys. Rev. Lett.}, 122:042503, 2019.

\bibitem{lei2019b}
J.~Lei and A.~M. Moro.
\newblock {\em Phys. Rev. Lett.}, 123:232501, 2019.

\bibitem{rangel2020}
J.~Rangel, M.R. Cortes, J.~Lubian, and L.~F. Canto.
\newblock {\em Phys. Lett. B}, 803:135337, 2020.

\bibitem{cortes2020}
M.~R. Cortes, J.~Rangel, J.~L. Ferreira, J.~Lubian, and L.~F. Canto.
\newblock {\em Phys. Rev. C}, 102:064628, 2020.

\bibitem{canto2015}
L.~F. Canto, P.~R.~S. Gomes, R.~Donangelo, J.~Lubian, and M.~S. Hussein.
\newblock {\em Phys. Rep.}, 596:1, 2015.

\bibitem{CH2008}
N.~Chamel and P.~Haensel.
\newblock {\em Living Rev. Rel.}, 11:10, 2008.

\bibitem{Steiner2012}
A.~W. Steiner.
\newblock {\em Phys. Rev. C}, 85:055804, 2012.

\bibitem{Loveland2007}
W.~Loveland.
\newblock {\em Phys. Rev. C}, 76:014612, 2007.

\bibitem{HM2000}
S.~Hofmann and G.~M\"unzenberg.
\newblock {\em Rev. Mod. Phys.}, 72:733, 2000.

\bibitem{HHO2013}
J.H. Hamilton, S.~Hofmann, and Y.T. Oganessian.
\newblock {\em Annual Review of Nuclear and Particle Science}, 63:383, 2013.

\bibitem{DHNO2015}
C.~E. Düllmann, {R.-D.} Herzberg, W.~Nazarewicz, and Y.~Oganessian.
\newblock {\em Nucl. Phys. A}, 944:1, 2015.
\newblock Special Issue on Superheavy Elements.

\bibitem{HDS2021}
D.J. Hinde, M.~Dasgupta, and E.C. Simpson.
\newblock {\em Prog. Part. Nucl. Phys.}, 118:103856, 2021.

\bibitem{Nazarewicz2018}
W.~Nazarewicz.
\newblock {\em Nature Physics}, 14:537, 2018.

\bibitem{Giuliani2019}
S.~A. Giuliani, Z.~Matheson, W.~Nazarewicz, E.~Olsen, P.-G. Reinhard,
  J.~Sadhukhan, B.~Schuetrumpf, N.~Schunck, and P.~Schwerdtfeger.
\newblock {\em Rev. Mod. Phys.}, 91:011001, 2019.

\bibitem{H2019}
K.~Hagino.
\newblock {\em AAPPS Bulletin}, 29:31, 2019.

\bibitem{myers1966}
W.~D. Myers and W.~J. Swiatecki.
\newblock {\em Nucl. Phys.}, 81:1, 1966.

\bibitem{sobiczewski1966}
A.~Sobiczewski, F.~A. Gareev, and B.~N. Kalinkin.
\newblock {\em Phys. Lett.}, 22:500, 1966.

\bibitem{koura2018}
H.~Koura, J.~Katakura, T.~Tachibana, and F.~Minato.
\newblock {\em Chart of the Nuclides}.
\newblock Japan Atomic Energy Agency, 2018.

\bibitem{bender1999}
M.~Bender, K.~Rutz, P.-G. Reinhard, J.~A. Maruhn, and W.~Greiner.
\newblock {\em Phys. Rev. C}, 60:034304, 1999.

\bibitem{oganessian04-2}
A.~Pakou, N.~Alamanos, G.~Doukelis, A.~Gillibert, G.~Kalyva, M.~Kokkoris,
  S.~Kossionides, A.~Lagoyannis, A.~Musumarra, C.~Papachristodoulou,
  N.~Patronis, G.~Perdikakis, D.~Pierroutsakou, E.~C. Pollacco, and K.~Rusek.
\newblock {\em Phys. Rev. C}, 69:054602, 2004.

\bibitem{kozulin2014}
E.~M. Kozulin, G.~N. Knyazheva, I.~M. Itkis, M.~G. Itkis, A.~A. Bogachev, E.~V.
  Chernysheva, L.~Krupa, F.~Hanappe, O.~Dorvaux, L.~Stuttg\'e, W.~H. Trzaska,
  C.~Schmitt, and G.~Chubarian.
\newblock {\em Phys. Rev. C}, 90:054608, 2014.

\bibitem{oganessian2000}
Yu.~Ts. Oganessian, V.~K. Utyonkov, Yu.~V. Lobanov, F.~Sh. Abdullin, A.~N.
  Polyakov, I.~V. Shirokovsky, Yu.~S. Tsyganov, G.~G. Gulbekian, S.~L.
  Bogomolov, B.~N. Gikal, A.~N. Mezentsev, S.~Iliev, V.~G. Subbotin, A.~M.
  Sukhov, O.~V. Ivanov, G.~V. Buklanov, K.~Subotic, M.~G. Itkis, K.~J. Moody,
  J.~F. Wild, N.~J. Stoyer, M.~A. Stoyer, R.~W. Lougheed, C.~A. Laue, Ye.~A.
  Karelin, and A.~N. Tatarinov.
\newblock {\em Phys. Rev. C}, 63:011301, 2000.

\bibitem{oganessian2004}
Yu.~Ts. Oganessian, V.~K. Utyonkov, Yu.~V. Lobanov, F.~Sh. Abdullin, A.~N.
  Polyakov, I.~V. Shirokovsky, Yu.~S. Tsyganov, G.~G. Gulbekian, S.~L.
  Bogomolov, B.~N. Gikal, A.~N. Mezentsev, S.~Iliev, V.~G. Subbotin, A.~M.
  Sukhov, A.~A. Voinov, G.~V. Buklanov, K.~Subotic, V.~I. Zagrebaev, M.~G.
  Itkis, J.~B. Patin, K.~J. Moody, J.~F. Wild, M.~A. Stoyer, N.~J. Stoyer,
  D.~A. Shaughnessy, J.~M. Kenneally, P.~A. Wilk, R.~W. Lougheed, R.~I.
  Il'kaev, and S.~P. Vesnovskii.
\newblock {\em Phys. Rev. C}, 70:064609, 2004.

\bibitem{Swiatecki2005}
W.~J. \ifmmode~\acute{S}\else \'{S}\fi{}wiatecki,
  K.~Siwek-Wilczy\ifmmode~\acute{n}\else \'{n}\fi{}ska, and
  J.~Wilczy\ifmmode~\acute{n}\else \'{n}\fi{}ski.
\newblock {\em Phys. Rev. C}, 71:014602, 2005.

\bibitem{ABGW2000}
Y.~Abe, D.~Boilley, {B. G.} Giraud, and T.~Wada.
\newblock {\em Phys. Rev. E}, 61:1125, 2000.

\bibitem{SKA2002}
C.~W. Shen, G.~Kosenko, and Y.~Abe.
\newblock {\em Phys. Rev. C}, 66:061602, 2002.

\bibitem{Swiatecki2003}
W.~J. \ifmmode~\acute{S}\else \'{S}\fi{}wiatecki,
  K.~Siwek-Wilczy\ifmmode~\acute{n}\else \'{n}\fi{}ska, and
  J.~Wilczy\ifmmode~\acute{n}\else \'{n}\fi{}ski.
\newblock {\em Acta Phys. Pol. B}, 34:2049, 2003.

\bibitem{AO2004}
Y.~Aritomo and M.~Ohta.
\newblock {\em Nucl. Phys. A}, 744:3, 2004.

\bibitem{ZG2015}
V.~I. Zagrebaev and W.~Greiner.
\newblock {\em Nucl. Phys. A}, 944:257, 2015.

\bibitem{KEWPIE2}
H.~L\"u, A.~Marchix, Y.~Abe, and D.~Boilley.
\newblock {\em Comput. Phys. Commun.}, 200:381, 2016.

\bibitem{Boilley16}
H.~L\"u, D.~Boilley, Y.~Abe, and C.~Shen.
\newblock {\em Phys. Rev. C}, 94:034616, 2016.

\bibitem{NRV}
\lq\lq{Nuclear Reaction Video"}.
\newblock http://nrv.jinr.ru/nrv/.

\bibitem{oganessian2013}
Yu.~Ts. Oganessian, V.~K. Utyonkov, F.~Sh. Abdullin, S.~N. Dmitriev,
  R.~Graeger, R.~A. Henderson, M.~G. Itkis, Yu.~V. Lobanov, A.~N. Mezentsev,
  K.~J. Moody, S.~L. Nelson, A.~N. Polyakov, M.~A. Ryabinin, R.~N. Sagaidak,
  D.~A. Shaughnessy, I.~V. Shirokovsky, M.~A. Stoyer, N.~J. Stoyer, V.~G.
  Subbotin, K.~Subotic, A.~M. Sukhov, Yu.~S. Tsyganov, A.~T\"urler, A.~A.
  Voinov, G.~K. Vostokin, P.~A. Wilk, and A.~Yakushev.
\newblock {\em Phys. Rev. C}, 87:034605, 2013.

\bibitem{yanez2014}
R.~Yanez, W.~Loveland, L.~Yao, J.~S. Barrett, S.~Zhu, B.~B. Back, T.~L. Khoo,
  M.~Alcorta, and M.~Albers.
\newblock {\em Phys. Rev. Lett.}, 112:152702, 2014.

\bibitem{Hinde1995}
D.~J. Hinde, M.~Dasgupta, J.~R. Leigh, J.~P. Lestone, J.~C. Mein, C.~R. Morton,
  J.~O. Newton, and H.~Timmers.
\newblock {\em Phys. Rev. Lett.}, 74:1295, 1995.

\bibitem{hagino2018}
K.~Hagino.
\newblock {\em Phys. Rev. C}, 98:014607, 2018.

\bibitem{Tanaka2018}
T.~Tanaka, Y.~Narikiyo, K.~Morita, K.~Fujita, D.~Kaji, K.~Morimoto, S.~Yamaki,
  Y.~Wakabayashi, K.~Tanaka, M.~Takeyama, A.~Yoneda, H.~Haba, Y.~Komori,
  S.~Yanou, {B. J.-P.} Gall, Z.~Asfari, H.~Faure, H.~Hasebe, M.~Huang,
  J.~Kanaya, M.~Murakami, A.~Yoshida, T.~Yamaguchi, F.~Tokanai, T.~Yoshida,
  S.~Yamamoto, Y.~Yamano, K.~Watanabe, S.~Ishizawa, M.~Asai, R.~Aono, S.~Goto,
  K.~Katori, and K.~Hagino.
\newblock {\em J. Phys. Soc. Japan}, 87:014201, 2018.

\bibitem{Hofmann2012}
S.~{Hofmann et al.}
\newblock {\em Euro. Pys. J. A}, 48:62, 2012.

\bibitem{Tanaka2020}
T.~Tanaka, K.~Morita, K.~Morimoto, D.~Kaji, H.~Haba, R.~A. Boll, N.~T. Brewer,
  S.~Van~Cleve, D.~J. Dean, S.~Ishizawa, Y.~Ito, Y.~Komori, K.~Nishio,
  T.~Niwase, B.~C. Rasco, J.~B. Roberto, K.~P. Rykaczewski, H.~Sakai, D.~W.
  Stracener, and K.~Hagino.
\newblock {\em Phys. Rev. Lett.}, 124:052502, 2020.

\end{thebibliography}
\bibliographystyle{unsrt}

\end{document}